\documentclass[twocolumn, nofootinbib]{aastex631}

\usepackage{amsmath,amssymb,amsfonts}          
\usepackage{xspace}                            
\usepackage{graphicx}                          
\usepackage{xcolor}                    
\usepackage[outdir=./auxiliary]{epstopdf}      
\graphicspath{{./}{auxiliary/}}

\newcommand{\pn}[1]{{\sc#1pn}\xspace}          
\newcommand{\into}{\rightarrow}                
\newcommand{\abs}[1]{\left\vert #1\right\vert} 

\newcommand{\del}[2]{\frac{\partial#1}{\partial #2}}     
\newcommand{\dif}[2]{\frac{d#1}{d#2}}                    

\newcommand{\bracket}[2]{\left\langle#1,#2\right\rangle} 
\newcommand{\dotprod}{\!\boldsymbol{\cdot}\!}                
\newcommand{\tcode}[1]{{\tt #1}}      					 
\newcommand{\m}[1]{\mathbf{#1}}         				 
\newcommand{\mxi }{\boldsymbol{\xi}}					 
\newcommand{\grad}{\boldsymbol{\nabla}} 				 

\newcommand{\BV}{Brunt-V\"ais\"al\"a }
\newcommand{\grpulse}{\tcode{GRPulse}}	

\defcitealias{ChriMull94}{JCD-DJM}
\defcitealias{PoisWill14}{P\&W}
\shortauthors{Boston, Evans, Clemens}
\shorttitle{Relativistic WD Asteroseismology}
\newcommand{\uncaddress}{Department of Physics and Astronomy, University of North Carolina, Chapel Hill, 
North Carolina 27599, USA}

\begin{document}

\title{Relativistic Corrections in White Dwarf Asteroseismology}

\correspondingauthor[0000-0001-8122-1961]{S.~Reece Boston}
\affiliation{\uncaddress}
\email{rboston@live.unc.edu}

\author[0000-0001-8122-1961]{S.~Reece Boston}
\affiliation{\uncaddress}

\author[0000-0001-5578-1033]{Charles R.~Evans}
\affiliation{\uncaddress}

\author[0000-0003-1964-2612]{J.~Christopher Clemens}
\affiliation{\uncaddress}

\begin{abstract}
With the precision now afforded by modern space-based photometric observations from the retired K2 
and current TESS missions, the effects of general relativity (GR) may be detectable in the light curves of 
pulsating white dwarfs (WDs).  Almost all WD models are calculated using a Newtonian description of gravity and 
hydrodynamics.  To determine if inclusion of GR leads to observable effects, we used idealized models of compact stars 
and made side-by-side comparison of mode periods computed using a
(i) Newtonian and 
(ii) GR description of the equilibrium structure and nonradial pulsations.  For application to white 
dwarfs, it is only necessary to include the first post-Newtonian (\pn1) approximation to GR.  The mathematical nature 
of the linear nonradial pulsation problem is then qualitatively unchanged and the GR corrections can be written as 
extensions of the classic Dziembowski equations.  As such, GR effects might easily be included in existing 
asteroseismology codes.  The idealized stellar models are 
(i) \pn1 relativistic polytropes and 
(ii) stars with cold degenerate-electron equation of state featuring a near-surface chemical transition 
from $\mu_e = 2$ to $\mu_e = 1$, simulating a surface hydrogen layer.  
Comparison of Newtonian and \pn1 normal mode periods reveals fractional 
differences on the order of the surface gravitational redshift $z$.  For a typical WD, this fractional difference is 
$\sim 10^{-4}$ and is greater than the period uncertainty $\sigma_{\Pi}/\Pi$ of many white dwarf pulsation 
modes observed by TESS.  A consistent theoretical modeling of periods observed in these stars should in principle 
include GR effects to \pn1 order. 
\end{abstract}



\section{Introduction}
\label{sec:intro}

With the exception of neutron stars, general relativity (GR) is not usually necessary for describing the equilibrium 
or dynamics of stars.  One caveat to this rule is that it has long been known \citep{HTWW65} that GR lowers the 
threshold for instability to radial collapse in compact stars at a local peak of a mass-radius relation.  For massive 
white dwarfs, this affects the exact value of the (idealized) Chandrasekhar mass $M_C$.  Relativistic polytropes 
provide a simple model for this effect of GR, where for a star of mass $M$ and radius $R$ the critical polytropic 
index $n_c$ for onset of instability is shifted \citep{ShapTeuk83} according to
\begin{equation}
\gamma_c = 1 + \frac{1}{n_c} = \frac{4}{3} + 2.25 \, \frac{GM}{c^2 R} .
\end{equation} 
The perturbation in critical exponent $\gamma_c$ is clearly proportional to the surface gravitational redshift 
$z = GM/(c^2 R)$, which serves as a post-Newtonian (PN) compactness parameter.  Recently, the effects of GR on 
ultra-massive white dwarfs have been considered \citep{MathNand17,CarvMariMalh18,NuneETC21,AlthETC22}.  GR effects 
on the internal structure become more significant as $M$ approaches $M_C$, the stellar radius shrinks, and the 
compactness parameter grows.

This paper is concerned with a different role played by GR, namely its effects on nonradial g-mode pulsations observed 
in ZZ Ceti and other variable WD stars.  It is reasonable to expect (which we confirm) that GR effects will show up 
at first order in the PN parameter (\pn1) in the conservative dynamics of g-modes.  While some variable WDs have 
higher mass, typical ZZ Ceti stars have $M \simeq 0.6 M_\odot$.  These stars have surface redshifts of order 
$z \simeq 10^{-4}$, leading to expected 
fractional shifts in g-mode periods at this level.  While these effects are obviously subtle, modern space-based 
photometry using TESS and K2 data, with long duration observations, yields fractional uncertainties in measuring 
g-mode periods down to $10^{-4}$ or better.  Some variables display modes (usually with periods in the range 
$\Pi \sim 100-200$s) that are coherent over lengths of time exceeding the spacecraft observations 
\citep[see][]{HermETC17B, HermETC13}.  Even longer duration time series, which include decades of ground-based 
observations, have revealed stars like G117-B15A \citep{KeplETC21} and R458 \citep{MukaETC13} with stable modes 
that have fractional period precisions better than $\sigma_\Pi/\Pi \sim 10^{-8}$, allowing attempts to measure WD 
cooling.

Recent white dwarf asteroseismology work \citep{GiamETC18} has attempted to build WD models that reproduce observed 
pulsation periods to the level of precision of space-based data.  Based on results of our calculations comparing 
Newtonian and \pn1 GR models, we argue that GR effects might sensibly be taken into account when trying to reproduce 
pulsation spectra with fractional line widths narrower than $10^{-4}$. 

The methods of asteroseismology (for main sequence stars as well as white dwarfs) have been reviewed in multiple 
settings \citep[e.g.][]{Cox80,HansKawa94,UnnoETC79}.  The Newtonian equations describing nonradial stellar 
pulsations are usually cast into a form due to \citet{Dzie71}, which introduces dimensionless variables and takes 
account of regular singular points in the system of equations at the stellar center (origin) and surface.  For example 
uses of these equations, see \citet{TownTeit13,TakaLoff04,CoriBenv02}.  Several community codes are available that 
calculate eigenmodes and periods once a stellar background model is specified.  These codes include \tcode{gyre} 
\citep[by][]{TownTeit13} and \tcode{adipls} \citep[by][]{Chri08}, which can be incorporated in stellar evolution 
codes, such as \tcode{MESA} \citep[by][]{PaxtETC10}.

The theory of stellar pulsations in GR developed largely separately, focused on the possibility of pulsations 
in neutron stars \citep{ThorCamp67,PricThor69,Thor69A,Thor69B,CampThor70,IpseThor73}.  For quadrupole or higher 
order ($\ell \ge 2$) nonradial modes, the gravitational field is dynamical and results in emission of gravitational 
waves (GWs).  The computed modes in these cases are quasinormal with complex eigenfrequencies that reflect damping.  
Later work \citep{LindDetw83, LindDetw85} simplified the system of perturbation equations to a fourth-order complex 
set, providing a means to more easily calculate modes numerically.  

It is not necessary to apply the full machinery of GR pulsation theory to WDs.  When pulsations of stars that are 
less compact than neutron stars are considered, like WDs, the conditions for applicability of both the slow-motion 
and weak-field approximations prevail, which means that the PN formalism can be used.  Applied to typical WDs, where 
the compactness parameter is $\sim 10^{-4}$, fractional corrections to mode periods will be of this order.  Even if 
\pn1 corrections can be observed, \pn2 is almost certainly unobservable.  Accordingly, we propose using the \pn1 
approximation as an accurate means of including GR effects in WD pulsation calculations \citep{Bost22} and have 
built an asteroseismology code, \grpulse, based on this premise.  For consistency, \pn1 GR would need to be included 
in both the pulsation equations and in the background stellar model.

It might be argued that the effects of GWs, which set in at \pn{2.5} order in the fluid motion, should be considered 
since they exhibit a secular behavior.  However, most of the observed WD g-modes are thought to be $\ell=1$ 
oscillations, which do not radiate GWs.  In those modes that do radiate, the secular effect would primarily manifest 
itself in the mode amplitude, and be much more 
difficult to measure accurately than mode periods.  Moreover, the $Q$ due to GW damping of an otherwise undisturbed 
free WD oscillation is enormous.  The coarsest approximation for the quadrupole f-mode would be 
$Q \sim (R c^2/G M)^{5/2} \sim 10^{10}$.  Additionally, for g-modes the $Q$ due to GW emission would be orders of 
magnitude larger still, since g-modes are primarily a motion near the surface with less mass involved in the 
perturbation $\delta\rho$ and (importantly) the periods, set by the Brunt-V\"ais\"al\"a frequency, are much longer 
than those of the f-mode or p-modes.  In any event, the observed sustained g-modes in WDs are not likely free 
oscillations but instead modes that are driven by the conversion of heat flux from the stellar interior into 
mechanical work, in the region of partial hydrogen ionization, by the trapping effects of the convection 
zone \citep{DoleVauc81,WingETC82,Bric91,GoldWu99} and damped by the onset of dissipative mode coupling parametric 
instability \citep{WuGold01}.  These stronger effects preclude the need to consider weaker GWs (which only apply 
to $\ell\ge 2$ pulsations), and our proposed use of the \pn1 approximation automatically avoids the issue.

The \pn1 gravitational field involves (i) a correction to the Newtonian potential $\Phi$, (ii) introduction of the 
gravitomagnetic vector potential $\m{W}$ (with only two nontrivial components, $W^r$ and $W^\theta$, in the case of 
fluid stars), and (iii) the presence of a new (\pn1) scalar potential $\Psi$.  Besides adding new (elliptic) field 
equations for these variables, numerous \pn1 correction terms appear in the fluid equations of motion.  As we 
show, in this format the usual linear pulsation equations can be extended in a transparent way.  

Stellar pulsations at \pn1 order have been considered previously \citep{Cutl91}, though in that paper applied to 
rotating neutron stars where the unperturbed stationary model is axisymmetric, not spherically symmetric.  In 
Newtonian terms, the rotational (centrifugal) potential is great enough to make the background model oblate.  We 
do not consider the complication of rapid rotation here.  

Nevertheless, it is worth remembering that some pulsating WDs do exhibit the effects of rotation.  Retaining only 
first power in the stellar angular velocity $\Omega$ (i.e., small rotation rate), the combined effects of the angular 
rate between the star and distant inertial observer and of the Coriolis accelerations $\m{\Omega} \times \delta\m{v}$ 
in the fluid motion, lead to a splitting of circular-mode periods proportional to their value of azimuthal mode number 
$m$ \citep{Cox80}.  These splittings are observed in the power spectra of WD light curves, with groupings of 
$2\ell + 1$ modes (e.g., triplets for $\ell=1$) allowing the mode number $\ell$ to be determined.  See 
\citet{KeplRobiNath83} for observed rotational splitting of $\ell=1$ in G226-29 and \citet{OdonWarn82} for splitting 
of $\ell=2$ and $\ell=1$ in L19-2.  Second-order-in-$\Omega$ effects of rotation on mode splittings are suggested by 
analysis of L19-2 \citep{OdonWarn82,BrasWeseFont89}.  Our calculations will be relevant for small rotation rates and 
when applied to the $m=0$ modes.

We note in passing that other approximations in mode calculations are sometimes made.  The most common is the 
Cowling approximation \citep{Cowl41} in which the perturbation in the gravitational potential is neglected.  One 
historical virtue of the Cowling approximation is it showed that the resulting second-order pulsation equations could 
be cast in nearly Sturm-Liouville form, and in the mode frequency limits $\omega \rightarrow \infty$ or 
$\omega \rightarrow 0$ the system assumes Sturm-Liouville form.  Out of this emerged the understanding of the 
existence of infinite sequences of p-modes and g-modes, separated by the single (for $\ell\ge2$) f-mode.  While the 
Cowling approximation of mode periods can be sufficiently accurate for some purposes, in this paper we seek to make 
an unambiguous comparison between Newtonian and \pn1 GR versions of pulsation theory, and therefore avoid the 
additional uncertainty associated with neglect of gravitational perturbations.  

In any event, while neglecting $\delta\Phi$ in the Newtonian system is straightforward enough, to include the Cowling 
approximation in our comparison would require a \pn1 version.  While a form of the Cowling approximation for the full 
GR pulsation theory has been developed by \citet{McDeVanHScho83} (see also \citep{McDeETC85,LindSpli90,YoshLee02}), 
the question arises of what parts of the metric perturbation to include and what to ignore \citep{Finn88}.  An 
advantage of the approximation is that it eliminates the radiative degree of freedom (GWs), yet this is true also 
when we adopt the \pn1 system.  

We note also that Finn has developed \citep{Finn86,Finn87} a slow-motion but not weak-field approximation for 
treating relativistic g-modes in neutron stars.  In our application to WDs, there is no need to try to retain the 
complexity of a strong field treatment.

To make comparison between Newtonian and \pn1 GR pulsations, we opt for simplified treatments of the background 
stellar model and adiabatic fluctuations in the perturbation equations.  We adopt two models for the star.  The 
first model treats the WD as a polytrope, with a particular choice of how to extend the Newtonian polytrope to GR.  
By picking an adiabatic index $\Gamma_1$ that is greater than the polytropic structural exponent $\gamma = 1 + 1/n$, 
the model supports g-mode oscillations.  While this affords a clean comparison, it has the disadvantage that 
g-modes in this model penetrate more deeply into the star and the resulting periods are much shorter than observed 
modes.  Accordingly, we have also generated a second model for the WD, in which the equation of state is just that 
of a cold degenerate electron gas, but one in which $\mu_e$ shifts smoothly from $\mu_e =2$ in the core to 
$\mu_e = 1$ in a surface layer.  This mimics the behavior of a surface hydrogen layer in a real WD.  In this model, 
the g-mode cavity is confined near the surface, and the mode periods more nearly approximate those seen in real 
variable WDs.  Neither of these stellar models is intended to be an accurate depiction of a real white dwarf.  
Instead, our goal is to keep the microphysical description simple in order to better highlight the differences 
in mode periods that occur depending upon whether we use a Newtonian or a \pn1 GR treatment.

This paper is organized as follows.  In Section \ref{sec:newtonian} we briefly review the standard Newtonian pulsation 
theory to set the notation.  In addition to Newtonian polytropes, we describe in Section \ref{subsec:0pnCHWD++} 
the stratified $T=0$ degenerate electron gas model that imitates a surface hydrogen layer.  In Section 
\ref{sec:1pn} we summarize the \pn1 formalism and introduce the linear wave equations at \pn1 order.  
This section also discusses (Sec.~\ref{sec:1pnPoly}) our particular choice for relativistic polytropes, and their 
\pn1 reduction, and (Sec.~\ref{sec:1pnCHWD++}) the \pn1 extension of our stratified $T=0$ models.  In Section 
\ref{sec:dziembowski} we reduce the Newtonian and \pn1 perturbation equations into a form more suitable for numerical 
study, which in the Newtonian case is due to \cite{Dzie71}.  Tests of the numerical behavior of our two versions of the 
code are summarized in Section \ref{sec:numerics}.  We make use, in particular, of tabulated periods of polytropes 
\citep{CutlLind92,LindMendIpse97} as code checks and in turn provide an expanded list for future reference.  Section 
\ref{sec:numerics} discusses the numerical performance of the code, which is designed to be more than adequate to 
accurately capture $\sim 10^{-4}$ fractional differences between Newtonian and \pn1 mode periods.  Section 
\ref{sec:results} gives our calculated numerical period shifts for both polytropic stars and the stratified models.  

\section{Newtonian Non-radial Pulsations}
\label{sec:newtonian}
For the sake of comparison to the \pn1 equations, we summarize linear adiabatic pulsations in Newtonian stars.
A fuller discussion is available in \cite{Cox80} and \cite{UnnoETC79}.  

\subsection{Newtonian Stellar Dynamics}
Consider a star with density $\rho(t,\m{r})$, pressure $P(t,\m{r})$, and Newtonian gravitational potential 
$\Phi(t,\m{r})$. The fluid motions inside the star are described by
\begin{subequations}\label{eq:0pnEOM}
\begin{align}
	\nabla^2\Phi &= 4\pi G \rho \label{eq:0pnEOMA}\\
	\partial_t \rho +\grad\dotprod\left( \rho \m{v}\right) &= 0\label{eq:0pnEOMB}\\
	\partial_t \left(\rho \m{v}\right) +\grad\dotprod \left(\rho \m{v}\m{v}\right)
				&= -\grad P -\rho\grad\Phi.\label{eq:0pnEOMC}
\end{align}
\end{subequations}
Here eqn.~\eqref{eq:0pnEOMB} is the familiar continuity equation, and eqn.~\eqref{eq:0pnEOMC} is Newton's 2nd Law
for a fluid under self-gravity.  In the static spherically-symmetric limit, eqn.~\eqref{eq:0pnEOMC} 
becomes the equation of hydrostatic equilibrium
\begin{equation}\label{eq:0pnstatic}
	0 = \dif{P}{r} + \rho \dif{\Phi}{r} .
\end{equation}

\subsection{Newtonian Polytropes}
We consider a simple model of a static star governed 
by a polytropic equation of state
\begin{equation}
\label{eq:PolytropeEOS}
	P = K \rho^{1+ \frac{1}{n}} ,
\end{equation}
where $n$ is the polytropic index.  When using this simple equation of state for a static, spherical star, equations 
\eqref{eq:0pnEOM} reduce to the Lane-Emden equation \citep{Lane1870} 
\begin{equation}\label{eq:0pnLE}
	\frac{1}{s^2}\dif{}{s}\left(s^2\dif{\theta}{s}\right) = -\theta^n ,
\end{equation}
where the Lane-Emden solution $\theta$ is related to the density and pressure by
\begin{equation}
\label{eq:Polytrope}
	\rho(r) = \rho_c \theta(s)^n, \qquad P(r) = P_c \theta(s)^{n+1} .
\end{equation}
The dimensionless radial variable $s$ is defined by%
\footnote{%
			In the Lane-Emden equation we use $s$ and not the common $\xi$ to 
			avoid confusion with $\delta\m{r}=\mxi$.}
\begin{equation}
\label{eq:PolytropeRadius}
	r = s \sqrt{\frac{(n+1)P_c}{4\pi G \rho_c^2}} .
\end{equation}
See e.g.~\citet{Chan39}.

\subsection{The Linearly Perturbed Newtonian Equations}
If an element of fluid is displaced by $\delta \m{r} = \mxi(t,\m{r})$, then the 
density, pressure and Newtonian potential will be perturbed in response.%
\footnote{%
			Throughout, we denote the Lagrange perturbations by $\delta$ 
			and Euler perturbations by $\Delta$.
			This is the opposite convention used in the GR literature,  
			e.g.~\cite{ShapTeuk83}. Our notation is
			a compromise with the Newtonian literature.}
The unperturbed background star is described by a solution to \eqref{eq:0pnstatic}.  For such a background, 
eqns.~\eqref{eq:0pnEOM} perturbed to first order become
\begin{subequations}
\label{eq:0pnPerturbed}
\begin{align}
	\nabla^2 \Delta\Phi &= 4 \pi G \Delta \rho \label{eq:0pnPerturbedA}\\
	\Delta\rho  &= -\grad\dotprod(\rho \mxi)\label{eq:0pnPerturbedB}\\
	\rho \frac{\partial^2 \mxi}{\partial t^2} &= -\grad\Delta P -\Delta\rho \grad\Phi 
			-\rho \grad\Delta\Phi.\label{eq:0pnPerturbedC}
\end{align}
\end{subequations}
These equations define a 4th-order system for the non-radial adiabatic perturbations.

In spherical stars, the normal modes can be decomposed into spherical harmonics.  Scalar perturbations 
become e.g., 
\begin{equation}
\Delta\rho(t,\m{r}) = \Delta\rho(r)_{\ell m} e^{i\omega t}Y_{\ell m}(\theta,\phi).
\end{equation} 
The displacement vector $\mxi$ 
is decomposed into a radial and a transverse vector spherical harmonic $\m{Y}_{\ell m} = r \grad Y_{\ell m}$,
\citep[see][]{BarrEsteGira85},
so that
\begin{equation}
\mxi(t,\m{r}) =  \xi_{\ell m}^r(r) \hat{\m{r}}Y_{\ell m} e^{i\omega t} + \xi_{\ell m}^H(r) \m{Y}_{\ell m} e^{i\omega t}.
\end{equation}
When broken up in this way, equations \eqref{eq:0pnPerturbed} describing the displacement can be written 
in component form,
\begin{subequations}
\label{eq:0pnPerturbedComponent}
\begin{align}
	\dif{\xi^r}{r} &= \left[\frac{g}{v_s^2} - \frac{2}{r}\right]\xi^r 
			+ \left[\frac{\ell(\ell+1)}{r^2\omega^2} - \frac{1}{v_s^2}\right]\chi 
			+ \frac{1}{v_s^2}\Delta\Phi,\\
	\dif{\chi}{r} &= [\omega^2 - N^2]\xi^r 
			+ \frac{N^2}{g}\chi 
			- \frac{N^2}{g}\Delta\Phi,
\end{align}
\end{subequations}
where $\chi=r\omega^2\xi^H$ and we suppress the $\ell,m$ indices on the perturbed quantities.

The quantity $N$ appearing in eqns.~\eqref{eq:0pnPerturbedComponent} is the \BV frequency, which along with the 
Lamb frequency \citep{Cox80} is important for classifying regions of mode stability.  The g-modes (gravity modes) 
are restored by buoyancy forces, and cannot propagate within convective regions in the star.  Stability against 
convection and propagation of g-modes depends on $N^2$ being positive \citep[see e.g.][]{KawaWingHans85}.
The \BV frequency is closely related to the Schwarzschild discriminant
\begin{equation}
	A = \dif{\log \rho}{r} - \frac{1}{\Gamma_1}\dif{\log P}{r} = - \frac{N^2}{g} ,
\end{equation}
where $g$ is the local gravity.  Thus, the Schwarzschild discriminant satisfies $A<0$ in regions where g-modes 
propagate.

In a spherically symmetric star, the g- and p-modes can be labeled by the angular momentum number $\ell$ (the modes 
are independent of azimuthal number $m$ in non-rotating stars), and further labeled by the principal radial mode 
number $k$ counting the radial nodes in each mode.  Using the Osaki-Scuflaire classification scheme 
\citep{Osak75,Scuf74}, we can classify $k<0$ as g-modes, $k>0$ as p-modes, and the fundamental mode (f-mode) has 
$k=0$. 

\subsection{Stratified Degenerate Electron Gas Models (CHWD++)}
\label{subsec:0pnCHWD++}
To begin building a simplified model that will mimic the pulsational behavior of WDs, we start with the 
Chandrasekhar WD (CHWD) equation of state, with the only pressure contribution coming from the completely degenerate 
($T=0$) electrons immersed in a sea of ions \citep{Chan39}.  We neglect electrostatic corrections.  The CHWD models 
are not immediately suitable for asteroseismology, since they have $N^2=0$ everywhere and do not admit g-modes 
solutions.  Real variable WDs have g-modes that are confined near the surface \citep{WingETC82}.  Important 
to explaining the presence of these near-surface modes is the addition of the Ledoux term in the 
\BV frequency \citep{BrasETC91}, associated with composition transitions occurring in the outer 
layers.  We can make a small adjustment to the CHWD $T=0$ equation of state by adding a composition transition, and 
thereby construct simple stellar models that support g-modes.

To account for a chemical transition, we consider $\mu_e$ (the mean molecular weight per electron) defined by
\begin{equation}
	\frac{1}{\mu_e} = \sum \frac{Z_i}{A_i} X_i ,
\end{equation}
where $Z_i$ and $A_i$ are the nuclear charge and nuclear mass numbers, respectively, and $X_i$ is the (local) mass 
fraction of any elemental species.  In the outermost layer composed of pure hydrogen, we will have $\mu_e = 1$, while 
$\mu_e=2$ in any region composed only of $\mbox{}^4$He, $\mbox{}^{12}$C, $\mbox{}^{16}$O, or combinations thereof.  
If an outer region exists where $X_{He}+X_H = 1$ and $X_{He}$ smoothly transitions from 1 to 0, then $\mu_e(r)$ will 
smoothly transition from 2 to 1, yielding the desired near-surface cavity.

\begin{figure*}
	\resizebox{\linewidth  }{!}{\input{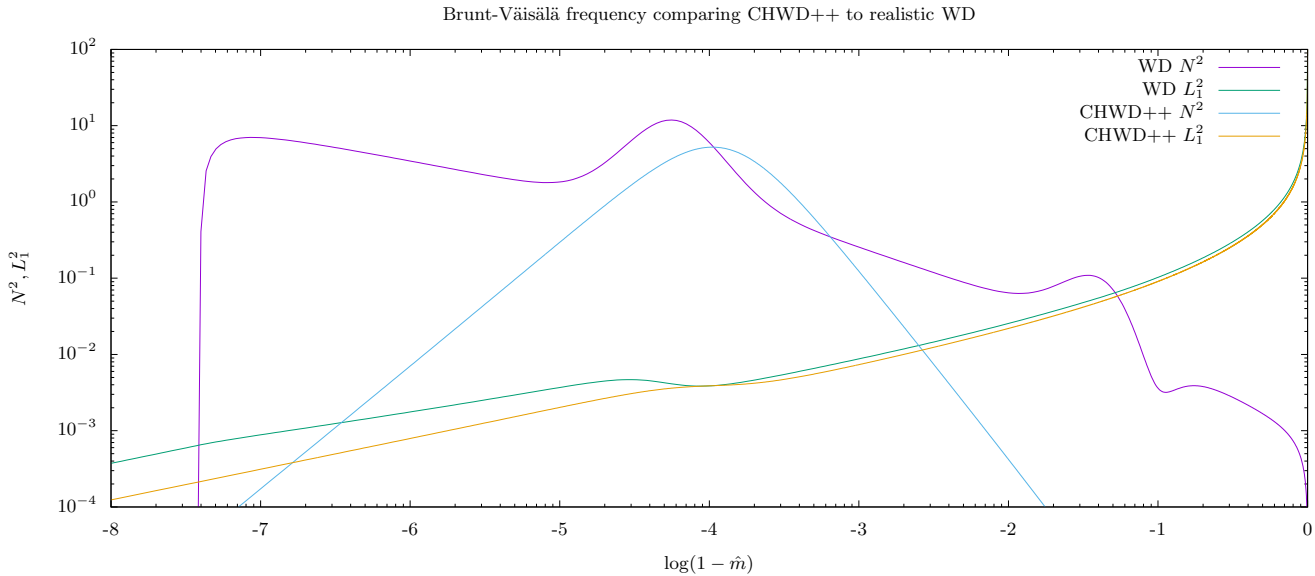}}
	\caption{%
		Comparison between the square of the \BV frequency $N^2$ found in the CHWD++ models (blue curve) 
		developed in this paper with $N^2$ that appears in realistic WD models (magenta curve) 
		\citep[see][]{Bost22}.  Also plotted is the Lamb frequency $L_1^2$ for $\ell=1$ (green and gold 
		curves) to highlight the presence of the g-mode cavities, which occurs in regions where the mode 
		frequency $\omega$ satisfies both $\omega^2 < N^2$ and $\omega^2 < L_1^2$.  The stellar surface is 
		located to the left in the plot and the stellar core to the right.  The presence of the composition 
		change in the outer layers allows for a non-zero $N^2$ in the stratified degenerate electron model, 
		forming a g-mode cavity qualitatively similar to those found in realistic WDs.
	\label{fig:CHWD}}
\end{figure*}

We take for the equation of state 
\begin{subequations}
\begin{align}
	\rho(r) &= B_0 \mu_e x^3 , \\
	P(r) &= A_0 f(x) ,
\end{align}
\end{subequations}
with $A_0$ and $B_0$ being parameters \citep{HansKawa94} dependent upon electron and proton masses and 
physical constants $h$ and $c$.  Here, $x$ and $f(x)$ are the dimensionless Fermi momentum and dimensionless 
degeneracy pressure function \citep{Chan39}, respectively, and $\mu_e$ is assumed to vary spatially.  To construct 
our stratified model, we set $\mu_e$ to be a function of pressure
\begin{equation}
\label{eq:0pnMue}
	\mu_e(r) = 1 + \frac{1}{1 + \exp\{\alpha(\log f_c-\log f(x))\}} .
\end{equation}
With appropriate choices for $\alpha$ and $f_c$, we can place the transition near the WD surface.  
Then in constructing the stellar model, $\mu_e$ will begin at $\mu_e(0) = 2$ in the center and near the surface 
smoothly transition to $\mu_e(R)=1$.  With this simple equation of state, the equation of hydrostatic equilibrium 
reduces (in terms of dimensionless radius $s$) to 
\begin{equation}
\label{eq:0pnCHWD++}
	\frac{1}{s^2} \dif{}{s}\left(s^2 \dif{y}{s}\right) 
			= -\mu_e^2(y^2-1)^{3/2} + \frac{1}{\mu_e} \dif{y}{s} \dif{\mu_e}{s},
\end{equation}
where $y = \sqrt{x^2-1}$ and $s = r B_0\sqrt{\pi G/2A_0}$.  Following \citet{BrasETC91} we can choose $\mu_e$ as the
unique indicator of composition change, and it can be shown that $N^2$ becomes
\begin{equation}
	N^2 =  -g\dif{\log \mu_e}{r} .
\end{equation}
Thus, g-modes will exist in any region of decreasing $\mu_e$, which we choose to be near the surface.
As seen in Fig.~\ref{fig:CHWD}, this leads to a curve for $N^2$ that has qualitative similarities to those seen 
in realistic WDs.  We refer to this cold degenerate but stratified model as the CHWD++ model \citep{Bost22}.  

\section{Post-Newtonian Non-radial Pulsations}
\label{sec:1pn}

Modern post-Newtonian theory, using either the MPM/PN (multipolar post-Minkowskian, post-Newtonian) approach or the 
DIRE (Direct Integration of the Relaxed Einstein equations) formulation, has been successful in self-consistently 
pushing PN results to high order.  See \citet{Blan14} and \citet{PoisWill14} for reviews.  For example, the source 
motion and gravitational waves in the two-body problem have been computed to third post-Newtonian order 
\citep{ArunETC08a} and beyond.  Our present interest in \pn1 is more modest and allows use of the simple classic 
approach to PN theory \citep{PoisWill14} (see their Section 8.2).  We briefly summarize the formalism here.

\subsection{Post-Newtonian Field Equations}
Gravity arises as the result of spacetime curvature responsing to the presence of matter (i.e., energy, momentum, and 
stress).  The gravitational field is described by the metric tensor $g_{\alpha\beta}$, which determines in part the 
line element that expresses infinitesimal proper distances and times between events.  At \pn1 order the line element 
can take the form
\begin{align}
\label{eq:1pnMetric}
\nonumber
	ds^2 &= g_{\alpha\beta}dx^\alpha dx^\beta \\
		&= -\left(1 + \frac{2\Phi}{c^2}+\frac{2(\Psi+\Phi^2)}{c^4}\right)c^2dt^2 
			+ \frac{8}{c^2} \vec{W}\dotprod d\vec{x} \, dt \nonumber\\
			&\qquad+ \left(1 - \frac{2\Phi}{c^2}\right) [dx^2+dy^2+dz^2] ,
\end{align}
where $\Phi$ is the usual Newtonian gravitational potential, $\m{W}$ is a \pn1 vector (gravitomagnetic) potential, 
and $\Psi$ is a \pn1 scalar potential.  (Much of the PN literature uses $U = -\Phi$ for the gravitational potential 
and opposite signs on the \pn1 fields.  Because we make connection with Newtonian asteroseismology, we retain the 
astrophysical convention for $\Phi$ and keep signs on $\m{W}$ and $\Psi$ consistent with that.)  The field is assumed 
to be weak, with the gravitational field spurring small deviations in the metric from that of flat space.  Factors 
of $1/c^2$ keep track of relative PN order.  The coordinates $x$,$y$,$z$ are nearly Minkowskian, and the harmonic 
gauge condition 
\begin{equation}
\label{eq:1pnHarmonic}
\partial_t \Phi + \partial_k W^k = 0 , 
\end{equation}
is adopted.  

The metric of eqn.~\eqref{eq:1pnMetric} is assumed to satisfy the Einstein field equations (EFE) 
(see \citealt{PoisWill14}, \citealt{Wein72})
\begin{equation}
\label{eq:EFE}
	G^{\alpha\beta} = \frac{8\pi G}{c^4} T^{\alpha\beta} ,
\end{equation}
term by term in the weak field and slow motion expansions.  Here $G^{\alpha\beta}$ is the Einstein tensor, which 
describes spacetime curvature, and $T^{\alpha\beta}$ is the stress-energy tensor, which describes the matter.  Both 
tensors depend on the metric $g_{\alpha\beta}$.  We take the matter to be a perfect fluid described by
\begin{equation}
T^{\alpha\beta} = \left(\rho + P/c^2 \right) u^{\alpha} u^{\beta} + P \,  g^{\alpha\beta} ,
\end{equation}
with $\rho$ the total energy density, $P$ the isotropic pressure, and $u^{\alpha}$ the fluid four-velocity.  The 
four-velocity is constrained by $u_{\alpha} u^{\alpha} = -c^2$, so that $u^{\alpha} = \gamma (c, v^k)$ with $\gamma$ 
determined from the coordinate velocity $v^k$ and the constraint.  We further define the baryon rest mass density 
$\rho_0$ that gives rise to baryon conservation $\nabla_{\alpha} \left(\rho_0 u^{\alpha}\right) = 0$.  Then $\rho$ 
and $\rho_0$ are related by defining the specific internal energy $\Pi$ such that $\rho = \rho_0 (1 + \Pi/c^2)$.  
The matter configuration is determined by a set of matter variables, such as $\{\rho, P, \Pi, v^k \}$.  
Alternatively, $\rho_0$ may be used in place of $\rho$.  Another more common alternative is to use the conserved, 
or rescaled, mass density $\rho^*$ \citep{PoisWill14} defined by $\rho^* = \sqrt{-g} \, \gamma \rho_0$.

Using the matter set $\{\rho, P, \Pi, v^k \}$ and recalling that in the slow-motion nearzone time derivatives 
of functions are of order $1/c$ compared to space derivatives, the \pn1 field equations become  
\begin{subequations}
\label{eq:1pnPotentials}
\begin{align}
	\label{eq:NewtonPoisson}
	\nabla^2\Phi &= 4\pi G \rho , \\
	\nabla^2\m{W} &= 4\pi G \rho\m{v} , \\
	\label{eq:PsiEqn}
	\nabla^2\Psi &= \del{^2\Phi}{t^2} + 4\pi G \left( 2 \rho v^2 -  2 \rho\Phi + 3 P \right) ,
\end{align}
\end{subequations}
where $\nabla^2$ is the flat-space Laplacian.  Leaving aside the choice of sign for the potentials, the use of $\rho$ 
in the fundamental matter set, rather than $\rho_0$ or $\rho^*$, affects the form of the source in \eqref{eq:PsiEqn} 
and therefore the definition of $\Psi$.  The form of the \pn1 field equations we list here is equivalent to that found 
in \citet{Wein72} but differs from that in \citet{PoisWill14}.  A more detailed derivation of these equations is found 
in \citet{Bost22}.  

\subsection{Post-Newtonian Hydrodynamics}

The \pn1 system is completed by deriving the \pn1 equations of motion of the perfect fluid.  The stress-energy tensor 
has vanishing covariant divergence
\begin{equation}
\label{eq:consvT}
	\nabla_\alpha T^{\alpha\beta}
	  = \partial_\alpha T^{\alpha\beta} + \Gamma^\beta_{ \alpha\gamma}T^{\alpha\gamma} 
	                                    + \Gamma^\alpha_{\alpha\gamma}T^{\gamma\beta} = 0 .
\end{equation}
Inserting the perfect fluid stress tensor and expansions for the connection, metric determinant, and four-velocity 
through \pn1 order, the time component of eqn.~\eqref{eq:consvT} is found to be
\label{eq:1pnEOM}
\begin{align}
 &\del{}{t}\left[\rho\left(1 + \frac{v^2-2\Phi}{c^2}\right)\right]
\label{eq:1pnEOMA}
\\
\nonumber
  &\quad + \grad\dotprod\left[\rho\m{v}\left(\!1 + \frac{v^2-2\Phi}{c^2}\!\right)\!+\!\frac{P\m{v}}{c^2}\right]
		 -\frac{\rho}{c^2}\del{\Phi}{t} = 0 ,
\end{align}
where $\grad$ is the flat-space operator.  The space components of eqn.~\eqref{eq:consvT} reduce to \pn1 order in a 
similar fashion
\begin{align}
 &\del{}{t} \left[\rho \m{v}\left( 1 + \frac{v^2-4\Phi}{c^2} \right) + \frac{P\m{v}}{c^2}\right]
\label{eq:1pnEOMB}
\\  \nonumber
	&\quad+ \grad\dotprod \left[\rho\m{v}\m{v}\left(1 + \frac{v^2-4\Phi}{c^2} \right) 
		+ \frac{P\m{v}\m{v}}{c^2}\right] + \grad P +\rho \grad \Phi 
\\  \nonumber
	&\quad+ \frac{1}{c^2}\bigg\{
			\rho\grad\Psi %
				+ (P + 2\rho v^2)\grad\Phi %
				+ 4\rho\del{\m{W}}{t}
\\  \nonumber 
	&\qquad
				- 2\rho \m{v} \del{\Phi}{t} %
				- 2\rho \m{v}(\m{v}\dotprod\grad\Phi) 
			 	- 4\rho\m{v}\times(\grad\times\m{W}) %
		\bigg\} = 0 .
\end{align}
Combining eqns.~\eqref{eq:1pnEOMA} and \eqref{eq:1pnEOMB} we can derive the \pn1 version of Euler's equation for the 
time rate of change of the (three) velocity.

For a relativistic barytropic equation of state, $P=P(\rho)$ (such as the \pn1 polytropes we consider below), 
eqns.~\eqref{eq:1pnPotentials} along with eqns.~\eqref{eq:1pnEOM} suffice to characterise the \pn1 system.  For a 
more general equation of state, where $P=P(\rho_0, \Pi)$, these equations must be supplemented with the expression 
for conservation of baryon number.  In the limit as $1/c \rightarrow 0$, everything reduces to the Newtonian system 
of eqns.~\eqref{eq:0pnEOM}.

In the static limit for a spherical configuration, eqn.~\eqref{eq:1pnEOMB} reduces to
\begin{equation}
\label{eq:1pnstatic}
	\dif{P}{r} + (\rho +P/c^2) \dif{\Phi}{r} + \frac{\rho}{c^2}\dif{\Psi}{r} = 0 , 
\end{equation}
which is the \pn1 version of the Tolman-Oppenheimer-Volkoff equation of relativistic hydrostatic equilibrium
\citep{Tolm39,OppeVolk39}.  

\subsection{\pn1 Polytropes}
\label{sec:1pnPoly}

For the case of a \pn1 static spherical polytrope, we follow Tooper's \citeyear{Toop64} solution to the 
equations of a static fluid sphere in GR \citep[see also][]{Tolm39,OppeVolk39}.  We specify an equation of state 
identical in form to eqn.~\eqref{eq:PolytropeEOS}, where $\rho = \epsilon/c^2$ represents the \emph{mass-energy 
density} inside the star (and not merely the baryon density as in Tooper's \citeyear{Toop65} paper).  We define the 
dimensionless relativistic parameter
\begin{equation}
\label{eq:1pnsigma}
	\sigma = \frac{P_c}{\rho_c c^2},
\end{equation}
specifying the strength of relativity within the star.  This $\sigma$ can be related to the surface redshift $z$.  
In addition to $\theta$ defined by eqn.~\eqref{eq:Polytrope} and $s$ defined by eqn.~\eqref{eq:PolytropeRadius}, 
define $\phi,\psi$ such that
\begin{subequations}
\begin{align}
	\Phi(r) &= \frac{(n+1)P_c}{\rho_c}\phi(s) , 
\\
	\Psi(r) &= \frac{(n+1)P_c^2}{\rho_c^2}\psi(s) .
\end{align}
\end{subequations}
In the static case eqns.~\eqref{eq:1pnPotentials}, \eqref{eq:1pnEOM} can be reduced to
\begin{subequations}
\label{eq:1pnLE}
\begin{align}
	\frac{1}{s^2}\dif{}{s}\left(s^2\dif{\theta}{s}\right) &= -\theta^n 
				\\&\quad + \sigma\left[2(n+1)\theta^n\phi - 4\theta^{n+1} 
				- \dif{\phi}{s}\dif{\theta}{s}\right] ,
\label{eq:1pnLEA}\nonumber\\
	\frac{1}{s^2}\dif{}{s}\left(s^2\dif{\phi}{s}\right) &= \theta^n , \\
	\frac{1}{s^2}\dif{}{s}\left(s^2\dif{\psi}{s}\right) &= 3\theta^{n+1}-2(n+1)\theta^n\phi ,
\end{align}
\end{subequations}
which are \pn1 Lane-Emden equations.  In this form it is clear that in the Newtonian limit $\sigma\into0$, 
$\dif{\phi}{s}\into-\dif{\theta}{s}$ and eqns.~\eqref{eq:1pnLE} reduce to eqns.~\eqref{eq:0pnLE}.  From 
eqn.~\eqref{eq:1pnstatic}, eqn.~\eqref{eq:1pnLEA} can be replaced by
\begin{equation}
	\dif{\theta}{s} = -(1+\sigma\theta)\dif{\phi}{s} - \sigma\dif{\psi}{s} ,
\end{equation}
which is better for numerical analysis.  The boundary conditions (BCs) on these equations are that $\theta=1$ at the 
center, $\theta=0$ at the surface, $d\phi/ds=d\psi/ds=0$ at the center, and $\phi$ must have an initial value so 
that $\Phi = -GM/R$ at the surface.  The value of $\sigma$ must be chosen to match the redshift by
\begin{equation}
	z = \left(1 - 2GM/c^2R\right)^{-1/2}-1 \approx (n+1)\sigma s_1\dif{\phi}{s_1} ,
\end{equation}
with $s_1$ the value of $s$ at the surface.  This matching of $\phi(0), \sigma$ must be accomplished by an iterative 
convergence routine.

\subsection{The Linearly Perturbed \pn1 Equations}
Consider a solution to eqns.~\eqref{eq:1pnEOM} for the case of a static, spherically symmetric star, and consider 
its response to a displacement of the fluid elements by $\delta\m{r} = \mxi$.  This will result in perturbations 
$\Delta\rho, \Delta P, \Delta \Phi, \Delta \Psi, \Delta \m{W}$.  Perturbed to first order on a static background,
eqns.~\eqref{eq:1pnEOM} will become
\begin{subequations}
\label{eq:1pnPerturbed}
\begin{align}
	\label{eq:1pnPertCont}
	0\!&=\!\grad\!\dotprod\!\left[ \mxi\!\left(\!\rho\!+\!\frac{P-2\rho\Phi}{c^2}\!\right) \!\right]%
		\!+\!\Delta\rho\left(\!\!1\!-\!\frac{2\Phi}{c^2}\!\right)%
		\!-\!\frac{3\rho\Delta\Phi}{c^2},\!\\
	0\!&=\!\left(\!\rho\!+\!\frac{P\!-\!4\rho\Phi}{c^2}\!\right)\!\del{^2\mxi}{t^2} \label{eq:1pnPertEuler1} %
		\!+\!\grad\!\Delta P\!+\!\rho\grad\Delta\Phi\!+\!\Delta\rho \grad\Phi %
		\\\nonumber&\quad 
			+ \frac{1}{c^2}\bigg\{
			\rho\grad\Delta\Psi %
			+ \Delta\rho \grad\Psi 
		\\\nonumber&\qquad\qquad%
			+ P\grad\Delta\Phi 
			+ \Delta P\grad\Phi 
			+ 4\rho\del{\Delta \m{W}}{t}\bigg\}.\nonumber %
\end{align}
\end{subequations}
The equations for the potentials are simply
\begin{subequations}
\begin{align}
	\nabla^2\Delta\Phi &= 4\pi G \Delta\rho , \\
	\nabla^2\Delta\m{W} &= 4\pi G \rho \partial_t \mxi , \\
	\nabla^2\Delta\Psi &= \partial_t^2 \Delta\Phi + 4\pi G[3 \Delta P - 2 \rho \Delta\Phi - 2\Phi\Delta\rho] .
\end{align}
\end{subequations}  
On a spherical background, the scalar perturbations of the normal modes will be proportional to a spherical harmonic 
$Y_{\ell m}$, and have harmonic time dependence $e^{i\omega t}$, analogously to the Newtonian case.  Because $\mxi$ 
and $\Delta\m{W}$ are vectors, they require vector spherical harmonics
\begin{subequations}
\begin{align}
\mxi(\m{r},t) &= \xi^r \hat{\m{r}} Y_{\ell m} e^{i\omega t} + \xi^H \m{Y}_{\ell m} e^{i\omega t} , \\
\Delta\m{W}(\m{r},t) &= \Delta W^r \hat{\m{r}} Y_{\ell m}e^{i\omega t} + \Delta W^H \m{Y}_{\ell m} e^{i\omega t} .
\end{align}
\end{subequations}
The vector Laplacian acting on $\Delta\m{W}$ will produce two second-order equations
\begin{subequations}
\begin{align}
	&\frac{1}{r^2}\dif{}{r}\left(r^2\dif{\Delta W^r}{r}\right)%
			-\frac{2+\ell(\ell+1)}{r^2}\Delta W^r 
\nonumber \\
&\qquad\qquad + \frac{2\ell(\ell+1)}{r^2}\Delta W^H = 4\pi G \rho i \omega \xi^r ,
\\
	&\frac{1}{r^2}\dif{}{r}\left(r^2\dif{\Delta W^H}{r}\right) 
				-\frac{\ell(\ell+1)}{r^2}\Delta W^H + \frac{2}{r^2}\Delta W^r \nonumber\\
		&\qquad\qquad= 4\pi G \rho i \omega\xi^H .
\end{align}
\end{subequations}  
This apparent 10th-order system actually reduces to the 8th-order system of eqn.~\eqref{eq:1pnSystem}, due to the 
harmonic coordinate condition of eqn.~\eqref{eq:1pnHarmonic}.  The perturbed version of the gauge condition is 
\begin{align}
\label{eq:1pnPertHarmonic}
0 &= \partial_t \Delta\Phi + \grad\dotprod\Delta\m{W}\\
&= i\omega \Delta\Phi(r) + \frac{1}{r^2}\dif{}{r}[r^2 \Delta W^r(r)] - \frac{\ell(\ell+1)}{r} \Delta W^H . \nonumber
\end{align}
This equation, and its derivative, allow us to eliminate two degrees of freedom from the system.

\subsection{\pn1 CHWD++}
\label{sec:1pnCHWD++}
The \pn1 generalization of the CHWD++ model introduced in \ref{subsec:0pnCHWD++} can be readily derived from the GR 
degenerate gas equation of state given by \citet{ChanToop64}, using their functions $g(x), h(x)$, with
\begin{subequations}
\begin{align}
	\rho(r) &= \rho_B(r) + \frac{1}{c^2} U_e(r) \nonumber\\
		&= B_0h(x,s)= B_0[ \mu_e(s) x^3 + \sigma g(x)] , \\
	P(r) &= A_0 f(x) ,
\end{align}
\end{subequations}
and with $\mu_e(x)$ as in \eqref{eq:0pnMue}.  Then the static equilibrium condition of \eqref{eq:1pnstatic} becomes
\begin{subequations}
\begin{align}
&\dif{x}{s} = - \frac{y}{x^4} \left\{ [h(x) + \sigma f(x)] \dif{\phi}{s} + \sigma h(x) \dif{\psi}{s}\right\} ,\\
&\dif{}{s} \dif{\phi}{s} = h(x)                - \frac{2}{s} \dif{\phi}{s} , \\
&\dif{}{s} \dif{\psi}{s} = 3f(x) - 16 h(x)\phi - \frac{2}{s} \dif{\psi}{s} ,
\end{align}
\end{subequations}
where $\Phi = 8A_0/B_0 \phi$ and $\Psi = 8A_0^2/B_0^2 \psi$.  The BCs are that $x=0$ at the surface, that 
$d\phi/ds=d\psi/ds=0$ at the center, and that $\phi$ must have an initial value so that $\Phi = -GM/R$ at the surface.  
The central value of $y$ (and hence $x$) is specified by the free parameter $y_0$, and $\sigma = A_0/B_0c^2$ is fixed 
by the relative masses of protons and electrons.

\section{The Dziembowski Form for Numerical Solution}
\label{sec:dziembowski}

We seek dimensionless forms of both eqn.~\eqref{eq:0pnPerturbed} and eqn.~\eqref{eq:1pnPerturbed} that are suitable 
for numerical study of the wave equations.  We choose the dimensionless form of \citet{Dzie71} for the Newtonian 
case, which we generalize to \pn1.

\subsection{The Newtonian Dziembowski Equations}
We begin with the Newtonian equations of \eqref{eq:0pnPerturbed}.  Define variables
\begin{align}
\label{eq:0pnDziembowski}
	y_1 = x^{2-\ell}\frac{\xi^r(r)}{r}, &\qquad
	y_2 = x^{2-\ell}\frac{\omega^2}{g} \xi^H(r),\\
	y_3 = x^{2-\ell}\frac{\Delta\Phi(r)}{gr}, &\qquad
	y_4 = x^{2-\ell}\frac{1}{g}\dif{\Delta\Phi(r)}{r},\nonumber
\end{align}
with $g=d\Phi/dr$, $x=r/R$, and the factor $x^{2-\ell}$ to improve behavior near the center.  We then define a set 
of dimensionless stellar structure quantities
\begin{align}
\label{eq:0pnstructure}
	A^\ast = \frac{1}{\Gamma_1}\dif{\log P}{\log r} - \dif{\log\rho}{\log r}, &\quad
	V_g = -\frac{1}{\Gamma_1}\dif{\log P}{\log r},\\
	U = \dif{\log g}{\log r}+2, &\quad
	c_1 = \frac{GM}{R^3}\frac{r}{g}, \nonumber
\end{align}
and define the dimensionless frequency
\begin{equation}\label{eq:omegabar}
	\bar{\omega}^2 = \omega^2 \frac{R^3}{GM} .
\end{equation}
The linear system describing the pulsations now assumes the Dziembowski form
\begin{subequations}
\label{eq:0pnSystem}
\begin{align}\label{eq:0pnSystem}
	x\dif{y_1}{x} &= \left[V_g - 3 + (2-\ell)\right] y_1 
\\ \nonumber&\qquad 
					+ \left[\frac{\ell(\ell+1)}{c_1\bar{\omega}^2} - V_g\right] y_2 + V_g y_3 ,
\\
	x\dif{y_2}{x} &= \left[c_1\bar{\omega}^2 - A^\ast\right]y_1 
\\ \nonumber&\qquad
					+ \left[1 + A^\ast - U + (2-\ell)\right] y_2 - A^\ast y_3 ,
\\
	x\dif{y_3}{x} &= \left[1-U + (2-\ell)\right]y_3 + y_4 ,
\\
	x\dif{y_4}{x} &= UA^\ast y_1 + UV_g y_2 + \left[\ell(\ell+1)-UV_g\right]y_3 
\nonumber
\\
&\qquad  +[(2-\ell)- U]y_4 .
\end{align}
\end{subequations}
Near the center, all solutions can be expanded in positive, even powers of $x$, and the central values satisfy
\label{eq:0pnBC}
\begin{equation}
\label{eq:0pnBC1}
	y_2(0) = \frac{c_1\bar{\omega}^2}{\ell} y_1(0), \qquad y_4(0) = \ell y_3(0).
\end{equation}
Near the surface, due to the vanishing pressure and density, the variables $A^\ast, V_g$ will both diverge.  
Nonetheless, imposing the condition $\delta P=0$ at the surface ($x=1$) leads to the BCs
\begin{equation}\label{eq:0pnBC2}
	y_2(1) = y_1(1) + y_3(1), \qquad y_4(1) = -(\ell+1) y_3(1) .
\end{equation}

While the BCs of eqns.~\eqref{eq:0pnBC} relate the boundary values, more care is needed for numerical solutions at 
both boundaries.  For improved accuracy, we use an even-powered series at the center \citep[see][Sec 17.6]{Cox80}, 
and follow the surface expansion approach detailed by \citet{ChriMull94} in their appendix.  See \citet{Bost22} for 
further details.

\subsection{The \pn1 Dziembowski Equations}
\label{subsubsec:dziem1PN}

We now consider the \pn1 equations of \eqref{eq:1pnPerturbed}.  We define variables $y_1$-$y_4$ analogously to 
eqn.~\eqref{eq:0pnDziembowski}, but replacing 
\begin{equation}
	g \rightarrow q = \dif{\Phi}{r}+\frac{1}{c^2}\dif{\Psi}{r} .
\end{equation}
We then define additional variables for the \pn1 potentials:
\begin{equation}
	z_1 = x^{2-\ell}\frac{\Delta\Psi}{qr\Phi_s} , \quad
	z_2 = x^{2-\ell}\frac{1}{q\Phi_s}\dif{\Delta\Psi}{r} ,
\end{equation}
and
\begin{equation}
	z_3 = x^{2-\ell}\frac{4i\omega }{q\Phi_s}\Delta W^r , \quad 
	z_4 = x^{2-\ell}\frac{4i\omega r}{q\Phi_s}\dif{\Delta W^r}{r} ,
\end{equation}
along with
\begin{equation}
	z_5 = x^{2-\ell}\frac{4i\omega }{q\Phi_s}\Delta W^H , \quad 
	z_6 = x^{2-\ell}\frac{4i\omega r}{q\Phi_s}\dif{\Delta W^H}{r} ,
\end{equation}
where $\Phi_s=-GM/R$ is the surface gravitational potential.  While there are ten variables, due to the harmonic 
coordinate condition only eight of them are independent.

As in the Newtonian case of eqn.~\eqref{eq:0pnstructure}, we define various background stellar quantities
\begin{subequations}
\label{eq:1pnstructure}
\begin{align}
	A^\ast &= \frac{1}{\Gamma_1}\dif{\log P}{\log r}\!-\!\frac{1}{\!1\!+\!P/\rho c^2\!}\dif{\log\rho}{\log r} ,
\\ \nonumber
	V_g    &=-\frac{1}{\Gamma_1}\dif{\log P}{\log r}, \;\;
	U      = \dif{\log q}{\log r} + 2, \;\;
	c_1    = \frac{GM}{R^3}\frac{r}{q} ,
\end{align}
\end{subequations}
and use $\bar{\omega}^2$ as in eqn.~\eqref{eq:omegabar}.  Additionally, we define
\begin{equation}
	\Phi^\ast = \Phi/\Phi_s, \quad \beta_\ast^2 = v_s^2/\Phi_s ,
\end{equation}
where $v_s = \sqrt{(\partial P/\partial\rho)_\mathrm{ad}}$ is the local sound speed.  To \pn1 order, the surface 
redshift is $z = \frac{GM}{c^2 R} = -\frac{\Phi_s}{c^2}$, which will be used as a relativistic PN compactness 
parameter.  

We can eliminate $z_4, z_6$ from all equations using eqn.~\eqref{eq:1pnPertHarmonic}, which leads to an 8th-order 
system of equations describing \pn1 pulsations:
\newpage
\begin{widetext}
\begin{subequations}
\label{eq:1pnSystem}
\begin{align}
	x\dif{y_1}{x} &=
		\left[V_g -3 -3zV_g\beta_\ast^2 + (2-\ell)\right]y_1
		+ \left[\frac{\ell(\ell+1)}{c_1\bar{\omega}^2} - V_g - 4 z V_g\Phi^\ast\right]y_2
		+ \left[ V_g -3zV_g\beta_\ast^2\right]y_3
		- zV_gz_1 - zV_g z_5 , \\
	x\dif{y_2}{x} &=
		\left[ c_1\bar{\omega}^2-A^\ast + z\left(4A^\ast\Phi^\ast-\frac{4 U\bar{\omega}^2 x^2}{\ell(\ell+1)}\right)\right]y_1
		+ \left[ 1+ A^\ast - U  - 4zV_g\beta_\ast^2 + (2-\ell)\right]y_2\\&\quad
		+ \left[ -A^\ast  + 4zA^\ast\Phi^\ast\right]y_3
		- z\left[\frac{4\bar{\omega}^2 x^2}{\ell(\ell+1)}\right]y_4
		 +z A^\ast z_1 + z A^\ast z_5 , \nonumber\\
	x\dif{y_3}{x} &= \left[1- U + (2-\ell)\right]y_3 + y_4 , \\
	x\dif{y_4}{x} &=
		 UA^\ast\left[1 + 2z\left(\frac{\beta_\ast^2}{\Gamma_1}-\Phi^\ast\right) \right]y_1
		+UV_g   \left[1 + 2z\left(\frac{\beta_\ast^2}{\Gamma_1}+\Phi^\ast\right) \right]y_2\\\nonumber &\quad
		+ \left[\ell(\ell+1)- UV_g\left(1 + 2z\left[\frac{\beta_\ast^2}{\Gamma_1}-\Phi^\ast\right]\right)\right]y_3
		+ [(2-\ell)- U] y_4
		+ zU V_g z_1
		+ zU V_g z_5 , \\
	x\dif{z_1}{x} &=\left[ 1 - U + (2-\ell)\right]z_1 + z_2 , \\ 
	x\dif{z_2}{x} &= 
		-2A^\ast U \Phi^\ast y_1
		+ \left[ 3UV_g\beta_\ast^2 - 2UV_g\Phi^\ast \right] y_2
		+\left[ 2UV_g\Phi^\ast- 5UV_g\beta_\ast^2 + \bar{\omega}^2 x^2\right]y_3 \\\nonumber&\quad
		+\ell(\ell+1)z_1 +[(2-\ell) - U] z_2 ,\\
	x\dif{z_3}{x} &= -4 \bar{\omega}^2 x^2 y_3 + [(2-\ell) - U] z_3 + \ell(\ell+1)z_5 , \\
	x\dif{z_5}{x} &= \frac{4U\bar{\omega}^2 x^2}{\ell(\ell+1)} y_1 + \frac{4 \bar{\omega}^2 x^2}{\ell(\ell+1)} y_4
						 + z_3 + [1 - U + (2-\ell)] z_5 .
\end{align}
\end{subequations}
\end{widetext}
We have derived equations \eqref{eq:1pnSystem} in two ways: one by using the classical PN approach directly; and 
again using the MPM/PN approach mentioned earlier.  Compare these equations to eqn.~\eqref{eq:0pnSystem}.  Notice 
that all new terms in these equations occur multiplied by the \pn1 compactness parameter $z$.  We therefore recover 
the classical Dziembowski equations in the Newtonian limit of $z \into 0$.

At the center of the star,
\begin{subequations}
\begin{align}
	y_2(0) = \frac{c_1\bar{\omega}^2}{\ell} y_1(0) &,\quad
	y_4(0) = \ell y_3(0) , \\
	z_2(0) = \ell z_1(0)&,\quad
	z_3(0) = \ell z_5(0) ,
\end{align}
\end{subequations}
while at the surface
\begin{subequations}
\begin{align}
	y_2(1) &= [y_1(1)+y_3(1)](1 - 3z\beta_\ast^2 - 4z\Phi^\ast)\\
		&\quad  - z[z_1(1)+z_5(1)] , \nonumber\\
	y_4(1) &= -(\ell+1) y_3(1) , \\
	z_2(1) &= -(\ell+1) z_1(1) , \\
	z_3(1) &= -(\ell+1) z_5(1) + \frac{4\bar{\omega}^2}{\ell} y_3(1) .
\end{align}
\end{subequations}
Similarly to the Newtonian problem, the solution at both the center and surface boundaries can be expanded in a 
power series, where the coefficients are determined only by the background model.

We draw a comparison between these \pn1 terms and the Newtonian Cowling approximation.  For nonradial pulsations, 
the full analysis should include matter perturbations, gravitational perturbations, and thermodynamic perturbations.  
Often, we are most interested in the adiabatic approximation, which includes only the matter and gravitational 
perturbations.  In the Cowling approximation, we further simplify by neglecting the gravitational perturbations and 
include only matter perturbations.  The adiabtic analysis can be considered to be the Cowling approximation with 
additional gravitational terms, and the nonadiabatic analysis considered to be the adiabatic approximation with 
additional thermodynamic terms.

In a similar way, the adiabatic \pn1 effects of this paper can be considered an additional add-on to the Newtonian 
adiabatic approximation, and can be handled similarly as the nonadiabatic effects (i.e. with numerical flags).  This 
makes the \pn1 equations \eqref{eq:1pnSystem} easy to integrate into codes written for Newtonian analysis.  This is 
another major benefit of the \pn1 approach.

In Section \ref{sec:results} we will discuss the eigenfrequencies of the system of equations \eqref{eq:1pnSystem} 
and compare with the Newtonian results.

\newpage
\section{Numerical Tests}
\label{sec:numerics}

To ensure parity in the analysis, we produced a code to calculate eigenmodes using Newtonian and \pn1 physics with 
polytropic backgrounds.  While there are published tables for frequencies in the Newtonian case 
\citep[such as][henceforth JCD-DJM]{ChriMull94} and readily available community codes (such as \tcode{gyre} and 
\tcode{adipls}), making our own Newtonian code grants an additional check to the \pn1 frequencies, that the 
differences are not due to method or machine.

\subsection{The \grpulse\ Asteroseismology Code}
Our code for both Newtonian and GR pulsations is called \href{https://zenodo.org/badge/latestdoi/442700026}{\grpulse}, 
originally introduced in \citet{Bost22}.  This code may be obtained from 
\href{https://github.com/rboston628/GRPulse}{GitHub}.  Documentation and sample input files to generate the tabulated 
values in this paper are available.  The program leverages object-oriented design for easy compatibility of different 
stellar models, different wave equations, or different integration methods.  We offer this code under the GNU General 
Public License.

The code is being further developed in two directions.  This present work highlights \grpulse's capabilities for 
Newtonian and \pn1 asteroseismology on simple models.  \grpulse\ is also capable of calculating frequencies in the 
GR Cowling approximation, and we anticipate further developing its abilities to the full GR mode equations of 
\citet{ThorCamp67,LindDetw85}.  We have also extended the range of Newtonian stellar models to include more realistic 
models of WDs beyond polytropes, which we will present in a future study.  We anticipate further expanding the number 
of models available, in each regime of physics.

\subsection{Polytropic Background Codes}
We calculate the Newtonian polytropic background by numerically solving eqn.~\eqref{eq:0pnLE} with simple RK4 on a 
uniform grid of fixed size $N_{star}$.  For $n=0, 1, 5$ there exist analytic solutions to test against 
\citep[see][Sec. 7.2]{HansKawa94}.  For a grid size $N_{star}=10^5$, we find relative errors from the analytic 
solutions always smaller than $10^{-13}$.  In addition, we can convert our solution to eqn.~\eqref{eq:0pnLE} 
in terms of $s,\theta$ to a solution in terms of physical variables such as $r, \rho, P, \Phi$ and insert 
these variables back into the original equations \eqref{eq:0pnEOM} to calculate a scaled residual, e.g., for 
\eqref{eq:0pnEOMA},
\begin{equation}
\label{eq:Residual}
 \text{res}(r) = \frac{\abs{ \dif{}{r}(r^2\dif{\Phi}{r})  - 4\pi G \rho r^2  }}
 		{\abs{\dif{}{r}(r^2\dif{\Phi}{r})}+\abs{4\pi G\rho r^2}} .
\end{equation}
Across a range of indices $n$, and for $N_{star}=10^5$, we find this residual to be on the order $10^{-12}$.  We can 
define a root-mean square residual (RMSR)
\begin{equation}
\label{eq:RMSR}
	\text{RMSR} = \sqrt{\frac{1}{R}\int_0^R \text{res}^2(r) dr} ,
\end{equation}
which gives an estimate of numerical error.  We can similarly define an RMSR for the eigenmodes, by defining an 
analog to eqn.~\eqref{eq:Residual} for eqn.~\eqref{eq:0pnPerturbed}.  If the RMSR is significantly smaller than the 
relative difference of the frequencies, we can be confident the difference is not due to numerical limitations in 
the calculation.

We calculate the \pn1 polytropic background using an identical method, but where the parameter $\sigma$ appearing in 
eqn.~\eqref{eq:1pnLE} must also be fixed to match the surface redshift $z = \frac{G M}{c^2 R}$.  There are no known 
solutions to the \pn1 polytrope equations, so no exact test can be performed.  Tests of the residuals in the original 
equations \eqref{eq:1pnEOM} are on the order $10^{-12}$ for $N_{star}=10^5$, across a range of $n$.

\begin{figure*}
	\resizebox{\linewidth  }{!}{\input{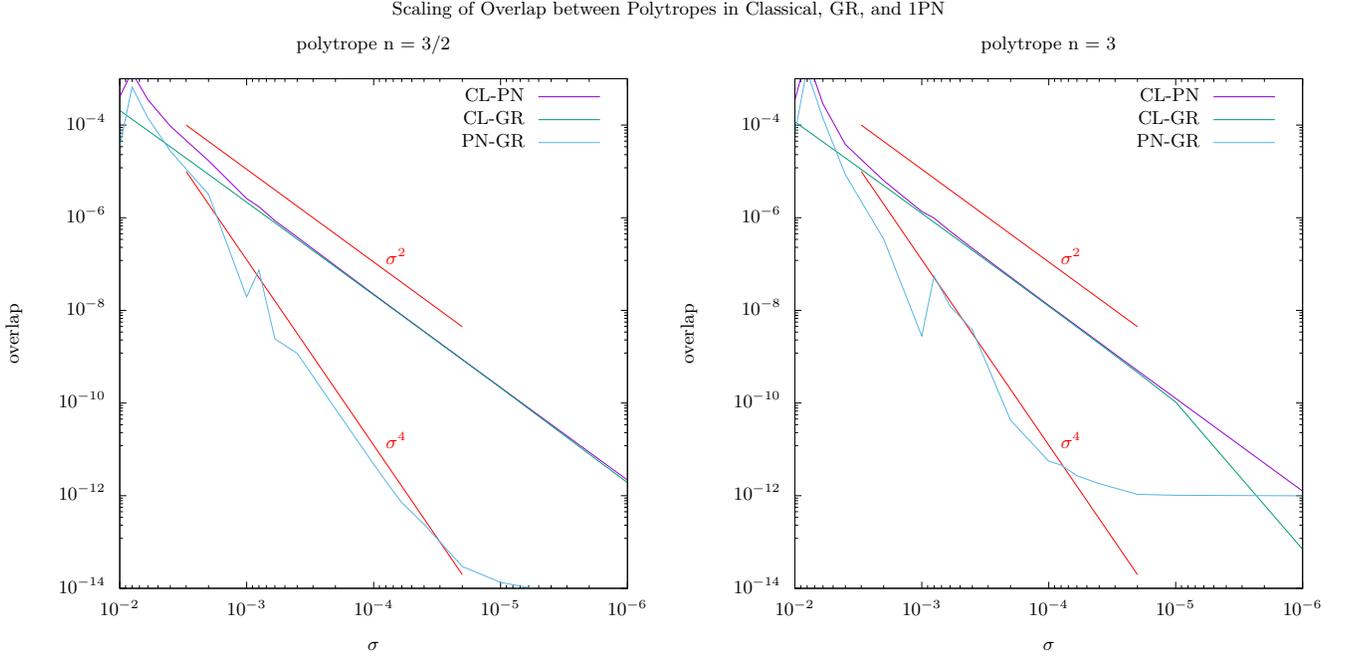}}
	\caption{Convergence of \pn1 and GR models to Newtonian as $\sigma\into0$.  
	The \pn1 and GR models converge to the Newtonian model like $\sigma^2$,
	but converge to each other like $\sigma^4$. Shown for $n=3/2$ (left) and $n=3$ (right).
	\label{fig:1pnPolytropeLimit}}
\end{figure*}

An additional test of the \pn1 polytrope is to calculate overlap coefficients with both the Newtonian (\pn0) and GR 
polytropic solutions, where the overlap is defined by
\begin{equation}
o(1,2) = 1 - \frac{\bracket{\theta_1}{\theta_2}}{\sqrt{\bracket{\theta_1}{\theta_1}\bracket{\theta_2}{\theta_2}}} ,
\end{equation}
with $\bracket{\theta_1}{\theta_2}$ being the usual inner product of functions.  The equation for a GR polytrope is 
a solution to the TOV equations with a polytropic equation of state, and has been explored in depth elsewhere 
\citep{Toop64,Toop65,Blud73}.  It is expected that $\theta_\text{\pn1}$ differs from $\theta_\text{\pn0}$ by an 
amount that scales with $\sigma$ defined in eqn.~\eqref{eq:1pnsigma}, and that therefore 
$o(\text{\pn0},\text{\pn1}) \sim \sigma^2$.  Similarly, $o(\text{\pn0},\text{GR}) \sim \sigma^2$, whereas 
$o(\text{\pn1},\text{GR})\sim \sigma^4$.  When both the \pn1 and GR solutions are matched to Schwarzschild 
coordinates and compared, we find the expected scaling in $\sigma$ for the overlaps, confirming that our \pn1 
polytrope accounts for GR up to order $\sigma$, with additional effects at order $\sigma^2$ (see Figure 
\ref{fig:1pnPolytropeLimit}).

\subsection{CHWD++ Background Codes}
We calculate the Newtonian and \pn1 CHWD++ background as above for the polytropes, using simple RK4 on a fixed grid 
of size $N_{star}$.  There are no exact solutions to compare against, though we can compare the mass-radius relation 
of these models to the mass-radius relationship of real WDs.  Tests of the residuals in \eqref{eq:0pnEOM} and 
\eqref{eq:1pnEOM} are on the order $10^{-12}$ for $N_{star}=10^5$.

\subsection{Newtonian Stellar Pulsation Code}
As with the background, we solve the pulsation equations using simple RK4 on a fixed, uniform grid.  We choose the 
grid for pulsations to be $N_{osc} = \tfrac{1}{2} N_{star}$ so that the calculated background values can be used in 
the half-steps of the RK4 method without interpolation.  The solution is found using a method similar to 
\citet{Chri08}.  We shoot from both the center and the surface to an internal fitting position $x_\mathrm{fit}$.  
At each boundary two independent solutions can be formed by choices of $y_1,y_3$ in eqns.~\eqref{eq:0pnBC1} and 
\eqref{eq:0pnBC2}.  The frequency $\bar{\omega}$ is adjusted to cause the two inward and two outward solutions to 
match, as determined by the vanishing of the Wronskian of the four solutions at $x_\mathrm{fit}$.  The physical 
solution is then made by a linear combination of the four solutions.

The resulting eigenmode is classified using the Osaki-Scuflaire method to identify mode order $k$ for p- and g-modes.  
This method counts the mode order by the zero-crossings on a graph of $y_1,y_2$, with clockwise crossings counted as 
negative.  Positive mode orders are considered p-modes, and negative are g-modes.

\begin{figure*}
	\resizebox{\columnwidth}{!}{\input{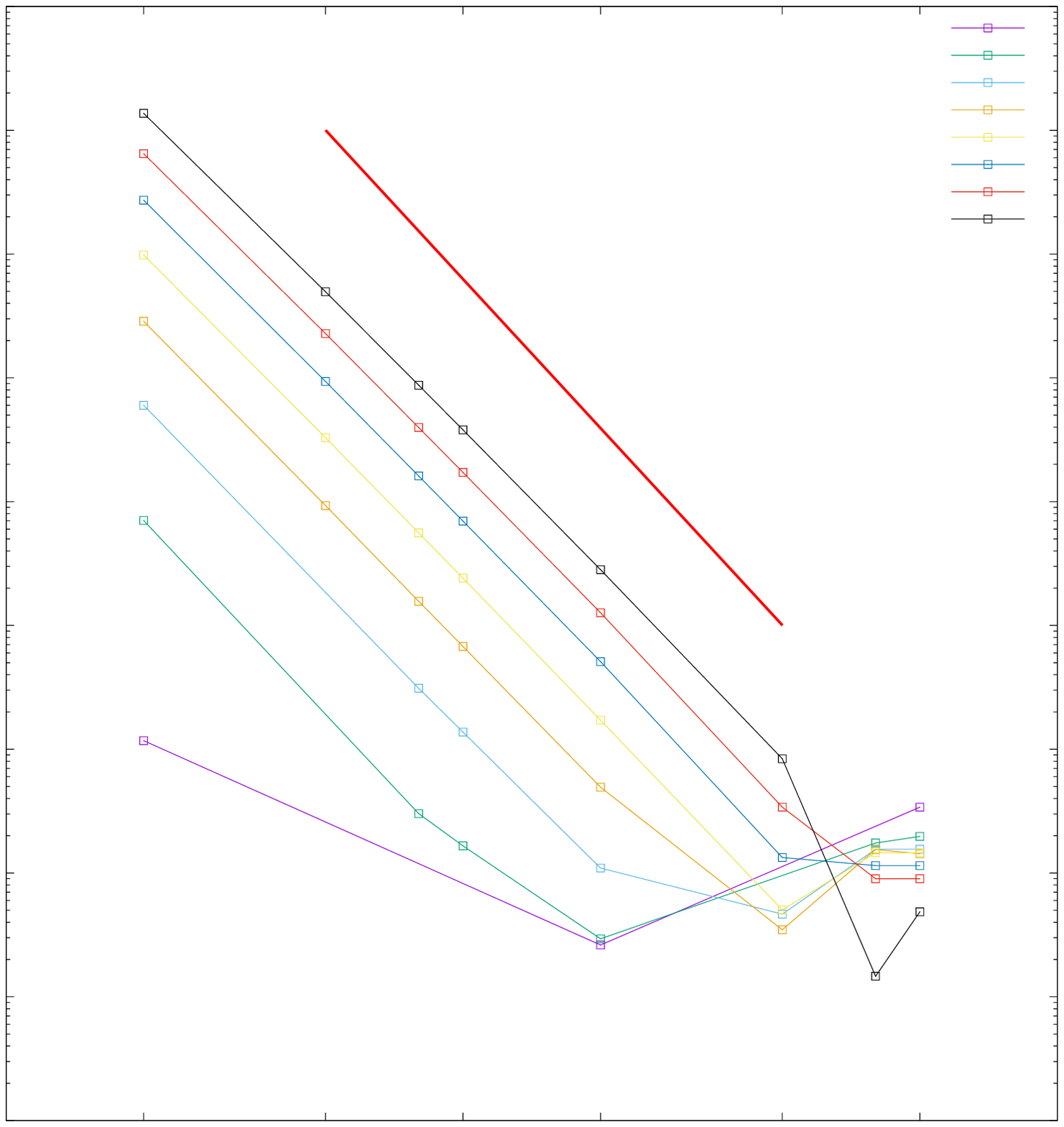}}
	\resizebox{\columnwidth}{!}{\input{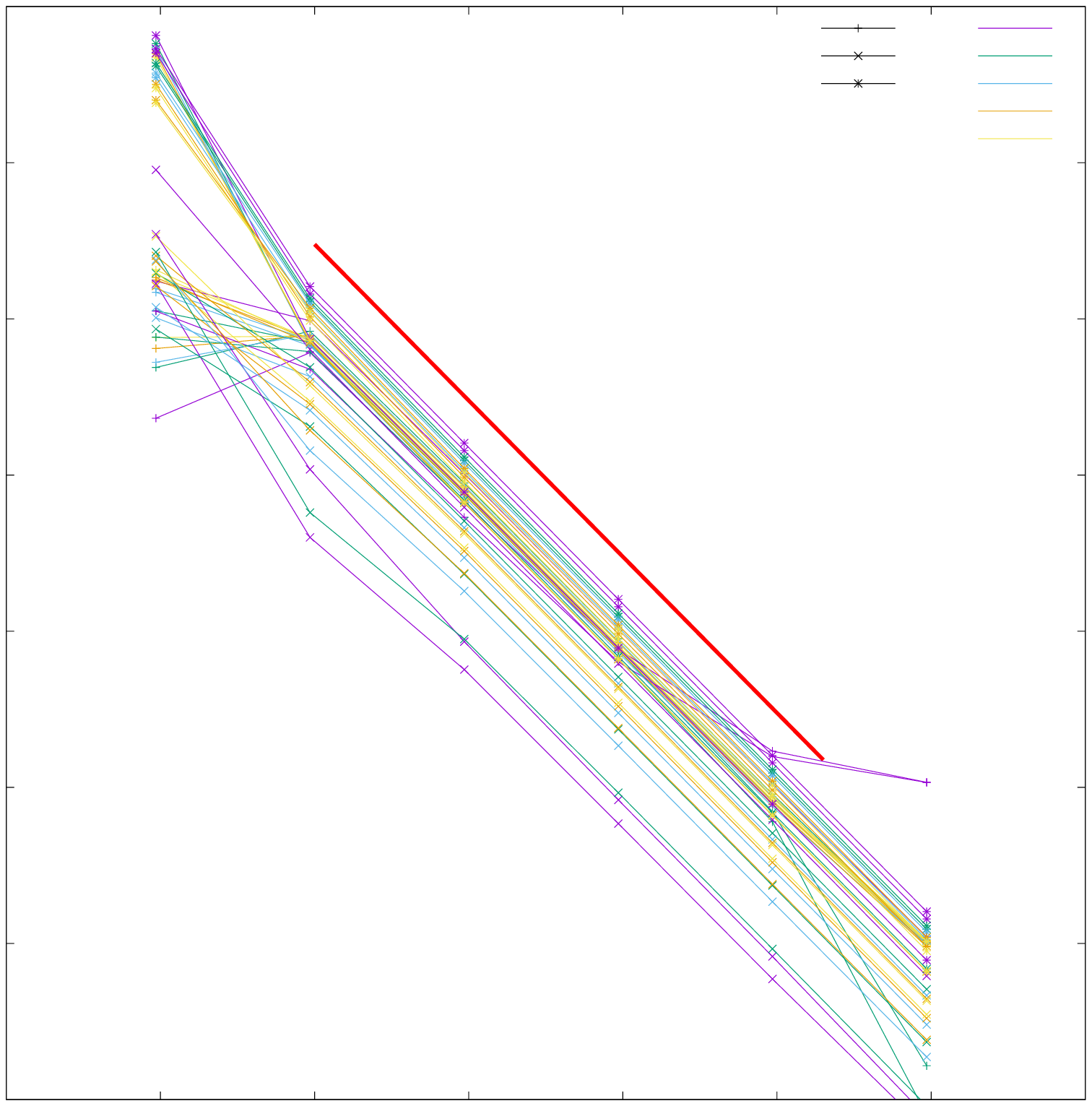}}
	\caption{
Left: The scaling of mode errors for $\ell=1$ modes calculated against eqn.~\eqref{eq:Pekeris} in an $n=0$ polytrope.  
	The errors scale like $N^{-4}$, and tends to increase for increasing $k$.  
	Right: The scaling of differences between Newtonian and \pn1 models for a range of redshifts $z$.  
	Several polytropes were used with $\ell=1-3$ and $k=1-5$.
	\label{fig:Pekeris}}
\end{figure*}

In the case of an $n=0$ polytrope (i.e., a uniform density star), there is an exact formula for the Newtonian
eigenfrequencies of p-modes due to \citet{Peke38}
\begin{equation}
\label{eq:Pekeris}
	\bar{\omega}^2_{k\ell} = D_{k\ell} + \sqrt{D_{k\ell}^2 + \ell(\ell+1)} , 
\end{equation}
where \citepalias[][eq.~3.3]{ChriMull94}
\begin{equation}
	D_{k\ell} = \Gamma_1 k(k+\ell+\tfrac{1}{2}) - 2 .
\end{equation}
Here $k=0,1,2,\ldots$ is the mode number, which counts radial nodes.  An equivalent form of this equation is found 
in \citet[][eq.~17.76]{Cox80}, where $n=0,1,2,\ldots$ is a recursion relation index, and is matched to mode number 
by $n=k-1$.  In the original of \citet[][eq.~32]{Peke38}, $n$ corresponds to $\ell$, and $k=0,2,4,\ldots$ is another 
recursion relation index, which corresponds to $2n$ as found in \citet{Cox80}.  The uniform density model allows
us to check the scaling of errors with $N_{star}$ and with the mode-order $k$.  Results are shown in Figure 
\ref{fig:Pekeris}.

We may also compare against the compiled tables of \citetalias{ChriMull94}, which are listed for p-modes to eight 
digits.  To make this comparison, we multiply $\bar{\omega}$ by their scaling factor $\nu_g= 99.855377\mu\text{Hz}$.  
When compared to JCD-DJM, we almost always find either no difference, or a difference of exactly $10^{-4}\mu$Hz 
(i.e., at the least significant digit).  The only exceptions occur due to an apparent difference in mode labeling 
in the low-order dipole ($\ell=1$) p-modes of the $n=4$ polytrope.  Using the Osaki-Scuflaire method, we were not 
able to find a $k=6$ mode.  The frequency we calculate for mode $k=5$ corresponds to what JCD-DJM call the $k=6$ mode, 
and this off-by-one problem filters down to modes $k=1,2,3,4$.  This is most likely due to the manual change to the 
low-order $\ell=1$ mode labels mentioned in the final paragraph in their appendix.

As with the background, the eigenfunctions, which are written in terms of the $y_i$ variables, can be converted into 
a solution in terms of $\Delta P, \m{\xi}$, etc., and inserted into eqn.~\eqref{eq:0pnPerturbed} to calculate an 
RMSR as in eqn.~\eqref{eq:RMSR}.  This residual is displayed in the tables in Section \ref{sec:results}.

\subsection{\pn1 Stellar Pulsation Code}
The same double-shooting RK4 method is used to solve the \pn1 pulsation equations.  Because there are 8 equations, 
there are two additional solutions at both the center and surface and the Wronskian is the determinant of an 
$8\times8$ matrix instead of a $4\times4$.  Otherwise, the \pn1 eigenmodes are found in an identical manner to the 
Newtonian case.  The eigenmode is also classified using the Osaki-Scuflaire method, again by counting crossings in 
$y_1, y_2$, defined as in the \pn1 approximation.

For the \pn1 case, the only error measurement we can use is the residual in the original physical equation.  Due to 
mathematical manipulations truncating at first-order in $z$, this residual scales with $z^2$, and for $z\sim10^{-4}$, 
the residual should be $\sim 10^{-8}$.  Because we are using identical numerical methods that we have verified in the 
Newtonian code, we can have additional confidence in the accuracy of the results.

There are no known analytic solutions for the \pn1 frequencies.  We do have a small table of f-mode frequencies to 
compare against, due to \citet[their table 1]{CutlLind92}, based on early work in the topic of \pn1 oscillations in 
rotating neutron stars.  They model the neutron star as an $n=1$ polytrope, and list frequencies for $\ell=2$ to 6 
calculated using $\Gamma_1=1+1/n$.  For the Newtonian star they use $M=1.736M_\odot$,$R=15.343$ km, and for the 
\pn1 star they use $M=1.4M_\odot$,$R=12.374$ km, which both give $z=0.2256$.  Their definition of the dimensionless 
frequency listed in their table 1 differs from $\bar{\omega}$ defined in eqn.~\eqref{eq:omegabar} by 
$\bar{\omega}_\text{CL} = \bar{\omega}\sqrt{4/3}$.  Accounting for this difference, we find our numbers compare to 
theirs as in Table \ref{tab:cutler}.  There is very close agreement on the Newtonian frequencies, and for the 
\pn1 frequencies differences are on the same order as the expected errors, $z^2\sim 0.05$, which account for 
methodological differences made in the \pn1 approximation.

There is a similar table in \citet[their table 1]{LindMendIpse97}, with $M=1.4M_\odot, R=14.45$ km, that lists both 
\pn1 frequencies and GR frequencies for the same star.  We find similar agreement with both sets of numbers, shown 
in Table \ref{tab:lindblom}.

The \pn1 approximation is not ideal for a highly compact object like a neutron star, unless numerical errors as 
high as 5\% are acceptable.  For a WD, the systemic errors in using the \pn1 approximation are orders of magnitude 
smaller.

\begin{table}
\caption{%
	Comparison of fundamental frequencies for corresponding Newtonian and \pn1 models with \citet{CutlLind92}.
\label{tab:cutler}}
\begin{center}
\vspace{-\baselineskip}
\begin{tabular}{l ccc B cccc}
\tableline\tableline
	& \multicolumn{3}{c}{Newtonian} 
	&& \multicolumn{4}{c}{Post-Newtonian, $z=0.2256$}\\
$\ell$	
	& $\bar{\omega}_{2022}$ & $\bar{\omega}_{1992}$ & \% err 
	&& $\bar{\omega}_{2022}$ & RMSR & $\bar{\omega}_{1992}$ & rel. dif.
	\\
	\tableline
$2$ 
	& 1.227 & 1.226 & 0.05\%      
	&& 1.317 & 0.07 & 1.232 & 0.06
	\\
$3$ 
	& 1.698 & 1.697 & 0.05\%      
	&& 1.694 & 0.06 & 1.606 & 0.05
	\\
$4$ 
	& 2.037 & 2.036 & 0.03\%     
	&& 1.984 & 0.05 & 1.885 & 0.05
	\\
$5$ 
	& 2.310 & 2.309 & 0.03\%    
	&& 2.228 & 0.05 & 2.120 & 0.05
	\\
$6$ 
	& 2.546 & 2.545 & 0.02\%     
	&& 2.444 & 0.04 & 2.324 & 0.05
	\\
\tableline
\end{tabular}
\end{center}
\vspace{-\baselineskip}
\tablecomments{%
Mass, radius as in their table 1.  All models are a polytrope with $n=1$ and 
$P\sim \rho^2 = (\epsilon/c)^2$, $\Gamma_1=1+1/n$.}
\end{table}

\begin{table}
\caption{Comparison of fundamental frequencies for corresponding \pn1 and GR models with \citet{LindMendIpse97}.
\label{tab:lindblom}}
\begin{center}
\vspace{-\baselineskip}
\begin{tabular}{l cc cc B cc}
\tableline\tableline
	& \multicolumn{4}{c}{Post-Newtonian, z=0.1836}
	&& \multicolumn{2}{c}{General Relativity}\\
$\ell$
	& $\bar{\omega}_{2022}$ & RMSR 
        & $\bar{\omega}_{1997}$ & rel. dif.  
        && $\bar{\omega}_{1997}$ &  rel. dif.\\
	\hline
$2$ & 1.279 & $0.05$ & 1.231 & 0.04 && 1.201 & 0.06\\
$3$ & 1.687 & $0.05$ & 1.619 & 0.04 && 1.586 & 0.06\\
$4$ & 1.989 & $0.04$ & 1.907 & 0.04 && 1.874 & 0.06\\
$5$ & 2.240 & $0.04$ & 2.147 & 0.04 && 2.113 & 0.06\\
\tableline
\end{tabular}
\end{center}
\vspace{-\baselineskip}
\tablecomments{%
Mass, radius as in their table 1.  All models are a polytrope with $n=1$ and 
$P\sim \rho^2 = (\epsilon/c)^2$, $\Gamma_1=1+1/n$.}
\end{table}

\section{Results}
\label{sec:results}

For the following calculations we model polytropes, scaled so that the total mass equals a typical field WD mass of 
$M = 0.6M_{\odot}$ (or $M=1.1934\times10^{33}$ g), with $R = 1.3 R_{\oplus}$ (or $R=8282$ km).  The radius is picked 
based on models of the WD mass-radius relationship.  There is an important subtlety here in how we choose background 
WD models in order to compare Newtonian and \pn1 mode periods.  We consider mass of the stars first.  For a 
Newtonian polytrope, the meaning of mass is clear.  When considering GR, the observed mass of a WD is the 
gravitational mass $M$, which is a combination of the integrated baryon mass, internal energy, and gravitational 
binding energy.  When we compare a Newtonian WD to a \pn1 WD, we keep the mass $M$ fixed.

For a Newtonian polytrope, the natural second parameter needed to specify the star is the radius, $R$, which is 
unambiguously defined.  Other quantities that are combinations of $M$ and $R$ might be used in place of $R$ in order 
to define the polytrope.  In GR, there is however a coordinate ambiguity in defining the radius of a star.  Our 
equations for the static background model are written in terms of an isotropic radial coordinate $r = r_\text{i}$, 
which appears when we convert the background line element \eqref{eq:1pnMetric} into spherical coordinates
\begin{align}
	ds^2 \!=\! &-\!\left(\!1\!+\!\frac{2\Phi}{c^2}\!+\!\frac{2(\Psi+\Phi^2)}{c^4}\!\right)\!c^2 dt^2 
\\ \nonumber&\qquad
	 + \left(\!1\!-\!\frac{2\Phi}{c^2}\!\right)\!\left(dr_\text{i}^2 + r_\text{i}^2  d\Omega^2\right) .
\end{align}
However, the radius of an isolated real WD can be indirectly estimated from observables like the surface gravitational 
redshift $z$ or the surface gravity $g_s$, determined by spectal line shifts and broadening, respectively.  For a 
spherical WD background, the exterior region, which connects the surface properties to a distant observer, is 
described by the Schwarzschild solution.  In standard Schwarzschild coordinates it has a line element given by
\begin{align}
	ds^2 = &-\left(1 - \frac{2 G M}{c^2 r_\text{a}}\right) c^2 \, dt^2 
\\ \nonumber&\qquad
	+ \left(1 - \frac{2 G M}{c^2 r_\text{a}} \right)^{-1}dr_\text{a}^2 + r_\text{a}^2 \, d\Omega^2 ,
\end{align}
where $r_\text{a}$ is the areal radial coordinate.  Any calculation of the radius from surface redshift 
$z = GM/(c^2 R)$ (or from the surface gravity) will give a value for $R = R_\text{a}$ in this latter coordinate 
system.  We make the choice that the radius we specify based on the mass-radius relation is the areal radius 
$R_\text{a}$.  On the other hand, the background star might be described in terms of isotropic coordinates, in 
which case its radius would be $R_\text{i}$, differing at \pn1 order from $R_\text{a}$.  These two radii can be 
related by
\begin{equation}
	R_\text{a} = R_\text{i} \sqrt{1 - \frac{2GM}{c^2R_\text{i}}} .
\end{equation}  
Given our choice for mass $M$ and radius $R_\text{a}$, we might alternatively view the surface redshift 
$z = GM/(c^2 R_\text{a}) = 1.07\times10^{-4}$ as the second polytropic parameter to hold fixed in comparing 
Newtonian and \pn1 models.  We adopt these values as parameters in our subsequent polytrope calculations.

In Tables \ref{tab:n=1.5} and \ref{tab:n=3}, we compile eigenfrequencies calculated on $n=1.5$ and $n=3$ polytrope 
backgrounds.  The table records the dimensionless frequency $\bar{\omega}$, the cyclic frequency $f$ (in Hz), the 
period $\Pi$ (in s), and the RMSR.  In an additional column, we specify the relative difference between $\bar{\omega}$ 
in the Newtonian and \pn1 calculations, defined in the sense
\begin{equation}
\label{eq:compare}
\text{rel. diff.} = \frac{\bar{\omega}_\text{\sc 1pn}-\bar{\omega}_\text{\sc 0pn}}{\bar{\omega}_{avg}} ,
\end{equation}
so that a positive relative difference means the \pn1 frequency is blueshifted, while a negative difference means 
redshifted.  The differences between periods are similar to those between $\bar{\omega}$.  The RMSR for the stars and 
modes is as defined in eqn.~\eqref{eq:RMSR}.  Additional tables for several values of $n$ can be found in 
\citep{Bost22}.

Note that g-modes are unstable for $n\leq1.5$, and that there is no f-mode ($k=0$) for $\ell=1$.

\begin{table*}
\begin{center}
\caption{Normal mode frequencies and periods for $n=1.5$ polytrope
\label{tab:n=1.5}}
\begin{tabular}{lrrrcrrrcc}
\tableline\tableline
	&\multicolumn{4}{c}{Newtonian Polytrope (RMSR = $2\times10^{-12}$)}
	&\multicolumn{4}{c}{Post-Newtonian Polytrope (RMSR = $2\times10^{-12}$)}
	& $z\sim10^{-4}$  \\
$l,k$
	& \multicolumn{1}{c}{$\bar{\omega}$}
	& \multicolumn{1}{c}{$f$ (Hz)}
	& \multicolumn{1}{c}{$\Pi$ (s)}
	& RMSR
	& \multicolumn{1}{c}{$\bar{\omega}$}
	& \multicolumn{1}{c}{$f$ (Hz)}
	& \multicolumn{1}{c}{$\Pi$ (s)}
	& RMSR
	& rel. diff. \\
\tableline
	1,1& 2.571761 & 0.153246 & 6.525466 & $4\times10^{-9}$& 2.571531 & 0.153207 & 6.527097 & $6\times10^{-8}$ & $-9.0\times10^{-5}$\\
	1,2& 4.256099 & 0.253612 & 3.943034 & $4\times10^{-9}$& 4.255778 & 0.253552 & 3.943964 & $7\times10^{-8}$ & $-7.5\times10^{-5}$\\
	1,3& 5.838061 & 0.347878 & 2.874575 & $4\times10^{-9}$& 5.837646 & 0.347797 & 2.875240 & $7\times10^{-8}$ & $-7.1\times10^{-5}$\\
	1,4& 7.373488 & 0.439370 & 2.275984 & $4\times10^{-9}$& 7.372978 & 0.439269 & 2.276507 & $7\times10^{-8}$ & $-6.9\times10^{-5}$\\
	1,5& 8.881992 & 0.529259 & 1.889435 & $4\times10^{-9}$& 8.881386 & 0.529138 & 1.889867 & $7\times10^{-8}$ & $-6.8\times10^{-5}$\\
	1,10& 16.248203 & 0.968196 & 1.032849 & $4\times10^{-9}$& 16.247119 & 0.967976 & 1.033084 & $7\times10^{-8}$ & $-6.7\times10^{-5}$\\
	1,15& 23.505846 & 1.400663 & 0.713948 & $4\times10^{-9}$& 23.504284 & 1.400345 & 0.714110 & $7\times10^{-8}$ & $-6.6\times10^{-5}$\\
	1,20& 30.727923 & 1.831011 & 0.546146 & $4\times10^{-9}$& 30.725885 & 1.830596 & 0.546270 & $7\times10^{-8}$ & $-6.6\times10^{-5}$\\
\tableline
	2,0& 1.455807 & 0.086748 & 11.527589 & $4\times10^{-3}$& 1.455893 & 0.086740 & 11.528757 & $6\times10^{-8}$ & $\hphantom{-}5.9\times10^{-5}$\\
	2,1& 3.207357 & 0.191120 & 5.232327 & $3\times10^{-9}$& 3.207218 & 0.191081 & 5.233395 & $6\times10^{-8}$ & $-4.3\times10^{-5}$\\
	2,2& 4.849223 & 0.288955 & 3.460748 & $4\times10^{-9}$& 4.848949 & 0.288892 & 3.461499 & $6\times10^{-8}$ & $-5.6\times10^{-5}$\\
	2,3& 6.426896 & 0.382965 & 2.611205 & $4\times10^{-9}$& 6.426508 & 0.382880 & 2.611782 & $7\times10^{-8}$ & $-6.0\times10^{-5}$\\
	2,4& 7.966983 & 0.474735 & 2.106436 & $4\times10^{-9}$& 7.966488 & 0.474630 & 2.106905 & $7\times10^{-8}$ & $-6.2\times10^{-5}$\\
	2,5& 9.482660 & 0.565051 & 1.769750 & $4\times10^{-9}$& 9.482060 & 0.564925 & 1.770146 & $7\times10^{-8}$ & $-6.3\times10^{-5}$\\
	2,10& 16.882254 & 1.005977 & 0.994058 & $4\times10^{-9}$& 16.881155 & 1.005750 & 0.994282 & $7\times10^{-8}$ & $-6.5\times10^{-5}$\\
	2,15& 24.160416 & 1.439668 & 0.694605 & $4\times10^{-9}$& 24.158830 & 1.439342 & 0.694762 & $7\times10^{-8}$ & $-6.6\times10^{-5}$\\
	2,20& 31.395293 & 1.870778 & 0.534537 & $4\times10^{-9}$& 31.393225 & 1.870355 & 0.534658 & $7\times10^{-8}$ & $-6.6\times10^{-5}$\\
\tableline
	3,0& 1.93432779 & 0.11526246 & 8.67585175 & $5\times10^{-3}$& 1.93436853 & 0.11524638 & 8.67706176 & $5\times10^{-8}$ & $\hphantom{-}2.1\times10^{-5}$\\
	3,1& 3.695765 & 0.220223 & 4.540858 & $3\times10^{-9}$& 3.695660 & 0.220181 & 4.541716 & $5\times10^{-8}$ & $-2.8\times10^{-5}$\\
	3,2& 5.348647 & 0.318714 & 3.137605 & $4\times10^{-9}$& 5.348409 & 0.318649 & 3.138248 & $6\times10^{-8}$ & $-4.5\times10^{-5}$\\
	3,3& 6.941521 & 0.413630 & 2.417617 & $4\times10^{-9}$& 6.941161 & 0.413543 & 2.418132 & $6\times10^{-8}$ & $-5.2\times10^{-5}$\\
	3,4& 8.496828 & 0.506308 & 1.975083 & $4\times10^{-9}$& 8.496354 & 0.506198 & 1.975510 & $6\times10^{-8}$ & $-5.6\times10^{-5}$\\
	3,5& 10.026632 & 0.597466 & 1.673737 & $4\times10^{-9}$& 10.026048 & 0.597335 & 1.674103 & $6\times10^{-8}$ & $-5.8\times10^{-5}$\\
	3,10& 17.477641 & 1.041455 & 0.960195 & $4\times10^{-9}$& 17.476538 & 1.041222 & 0.96042- & $7\times10^{-8}$ & $-6.3\times10^{-5}$\\
	3,15& 24.785387 & 1.476908 & 0.677090 & $4\times10^{-9}$& 24.783786 & 1.476576 & 0.677243 & $7\times10^{-8}$ & $-6.5\times10^{-5}$\\
	3,20& 32.038738 & 1.909120 & 0.523802 & $4\times10^{-9}$& 32.036649 & 1.908689 & 0.523920 & $7\times10^{-8}$ & $-6.5\times10^{-5}$\\
\tableline
\end{tabular}
\end{center}
\tablecomments{%
Both stars are scaled to $M=0.6M_{\odot}$ and $R=1.3R_{\oplus}$. %
All modes use $\Gamma_1 = 5/3$.  %
There are no g-modes for this model.}
\end{table*}

\begin{table*}
\begin{center}
\caption{Normal mode frequencies and periods for $n=3$ polytrope
\label{tab:n=3}}
\begin{tabular}{lrrrcrrrcc}
\tableline\tableline
	&\multicolumn{4}{c}{Newtonian Polytrope (RMSR = $2\times10^{-12}$)}
	&\multicolumn{4}{c}{Post-Newtonian Polytrope (RMSR = $2\times10^{-12}$)}
	& $z\sim10^{-4}$  \\
$l,k$
	& \multicolumn{1}{c}{$\bar{\omega}$}
	& \multicolumn{1}{c}{$f$ (Hz)}
	& \multicolumn{1}{c}{$\Pi$ (s)}
	& RMSR
	& \multicolumn{1}{c}{$\bar{\omega}$}
	& \multicolumn{1}{c}{$f$ (Hz)}
	& \multicolumn{1}{c}{$\Pi$ (s)}
	& RMSR
	& rel. diff. \\
\tableline\tableline
	1,-10& 0.343611 & 0.020475 & 48.839961 & $9\times10^{-12}$& 0.343666 & 0.020475 & 48.839927 & $4\times10^{-7}$ & $1.6\times10^{-4}$\\
	1,-5& 0.608215 & 0.036242 & 27.592119 & $3\times10^{-12}$& 0.608315 & 0.036242 & 27.592025 & $4\times10^{-7}$ & $1.6\times10^{-4}$\\
	1,-4& 0.719567 & 0.042877 & 23.322293 & $3\times10^{-12}$& 0.719686 & 0.042878 & 23.322150 & $4\times10^{-7}$ & $1.7\times10^{-4}$\\
	1,-3& 0.880757 & 0.052482 & 19.053998 & $2\times10^{-12}$& 0.880909 & 0.052483 & 19.053777 & $3\times10^{-7}$ & $1.7\times10^{-4}$\\
	1,-2& 1.133891 & 0.067566 & 14.800318 & $2\times10^{-12}$& 1.134099 & 0.067568 & 14.799968 & $3\times10^{-7}$ & $1.8\times10^{-4}$\\
	1,-1& 1.586168 & 0.094516 & 10.580181 & $1\times10^{-12}$& 1.586506 & 0.094521 & 10.579625 & $2\times10^{-7}$ & $2.1\times10^{-4}$\\
\tableline
	1,1& 3.377036 & 0.201230 & 4.969429 & $1\times10^{-12}$& 3.377300 & 0.201214 & 4.969838 & $2\times10^{-7}$ & $7.8\times10^{-5}$\\
	1,2& 4.642432 & 0.276633 & 3.614903 & $2\times10^{-12}$& 4.642751 & 0.276607 & 3.615235 & $2\times10^{-7}$ & $6.9\times10^{-5}$\\
	1,3& 5.909240 & 0.352119 & 2.839949 & $2\times10^{-12}$& 5.909632 & 0.352086 & 2.840217 & $2\times10^{-7}$ & $6.6\times10^{-5}$\\
	1,4& 7.176668 & 0.427642 & 2.338403 & $2\times10^{-12}$& 7.177142 & 0.427602 & 2.338624 & $2\times10^{-7}$ & $6.6\times10^{-5}$\\
	1,5& 8.443277 & 0.503117 & 1.987610 & $2\times10^{-12}$& 8.443837 & 0.503069 & 1.987797 & $2\times10^{-7}$ & $6.6\times10^{-5}$\\
	1,10& 14.751133 & 0.878988 & 1.137671 & $1\times10^{-11}$& 14.752152 & 0.878908 & 1.137775 & $2\times10^{-7}$ & $6.9\times10^{-5}$\\
\tableline\tableline
	2,-10& 0.567887 & 0.033839 & 29.551565 & $2\times10^{-11}$& 0.567978 & 0.033839 & 29.551593 & $4\times10^{-7}$ & $1.6\times10^{-4}$\\
	2,-5& 0.967663 & 0.057661 & 17.342746 & $4\times10^{-12}$& 0.967817 & 0.057661 & 17.342780 & $4\times10^{-7}$ & $1.6\times10^{-4}$\\
	2,-4& 1.127173 & 0.067166 & 14.888523 & $4\times10^{-12}$& 1.127352 & 0.067166 & 14.888544 & $4\times10^{-7}$ & $1.6\times10^{-4}$\\
	2,-3& 1.349915 & 0.080439 & 12.431848 & $2\times10^{-12}$& 1.350133 & 0.080439 & 12.431843 & $3\times10^{-7}$ & $1.6\times10^{-4}$\\
	2,-2& 1.681711 & 0.100210 & 9.979088 & $2\times10^{-12}$& 1.681991 & 0.100210 & 9.979029 & $3\times10^{-7}$ & $1.7\times10^{-4}$\\
	2,-1& 2.216884 & 0.132099 & 7.570059 & $2\times10^{-12}$& 2.217291 & 0.132102 & 7.569883 & $3\times10^{-7}$ & $1.8\times10^{-4}$\\
\tableline
	2,0& 2.859255 & 0.170377 & 5.869340 & $1\times10^{-12}$& 2.859867 & 0.170386 & 5.869026 & $2\times10^{-7}$ & $2.1\times10^{-4}$\\
\tableline
	2,1& 3.906874 & 0.232802 & 4.295491 & $1\times10^{-12}$& 3.907499 & 0.232802 & 4.295493 & $2\times10^{-7}$ & $1.6\times10^{-4}$\\
	2,2& 5.169469 & 0.308038 & 3.246357 & $1\times10^{-12}$& 5.170107 & 0.308026 & 3.246478 & $2\times10^{-7}$ & $1.2\times10^{-4}$\\
	2,3& 6.439990 & 0.383745 & 2.605895 & $2\times10^{-12}$& 6.440673 & 0.383724 & 2.606037 & $2\times10^{-7}$ & $1.1\times10^{-4}$\\
	2,4& 7.708951 & 0.459360 & 2.176942 & $2\times10^{-12}$& 7.709691 & 0.459330 & 2.177083 & $2\times10^{-7}$ & $9.6\times10^{-5}$\\
	2,5& 8.975891 & 0.534854 & 1.869668 & $2\times10^{-12}$& 8.976697 & 0.534816 & 1.869802 & $2\times10^{-7}$ & $9.0\times10^{-5}$\\
	2,10& 15.284901 & 0.910795 & 1.097942 & $2\times10^{-11}$& 15.286091 & 0.910719 & 1.098033 & $2\times10^{-7}$ & $7.8\times10^{-5}$\\
\tableline\tableline
	3,-10& 0.766497 & 0.045674 & 21.894321 & $2\times10^{-11}$& 0.766618 & 0.045674 & 21.894388 & $4\times10^{-7}$ & $1.6\times10^{-4}$\\
	3,-5& 1.259737 & 0.075065 & 13.321780 & $5\times10^{-12}$& 1.259929 & 0.075064 & 13.321887 & $4\times10^{-7}$ & $1.5\times10^{-4}$\\
	3,-4& 1.446622 & 0.086201 & 11.600779 & $9\times10^{-12}$& 1.446840 & 0.086200 & 11.600890 & $4\times10^{-7}$ & $1.5\times10^{-4}$\\
	3,-3& 1.699020 & 0.101241 & 9.877421 & $4\times10^{-12}$& 1.699274 & 0.101240 & 9.877531 & $3\times10^{-7}$ & $1.5\times10^{-4}$\\
	3,-2& 2.058262 & 0.122647 & 8.153451 & $2\times10^{-12}$& 2.058568 & 0.122646 & 8.153550 & $3\times10^{-7}$ & $1.5\times10^{-4}$\\
	3,-1& 2.601340 & 0.155008 & 6.451267 & $2\times10^{-12}$& 2.601732 & 0.155007 & 6.451330 & $3\times10^{-7}$ & $1.5\times10^{-4}$\\
\tableline
	3,0& 3.068190 & 0.182827 & 5.469654 & $2\times10^{-12}$& 3.068607 & 0.182822 & 5.469790 & $3\times10^{-7}$ & $1.4\times10^{-4}$\\
\tableline
	3,1& 4.294602 & 0.255906 & 3.907682 & $2\times10^{-12}$& 4.295218 & 0.255902 & 3.907750 & $2\times10^{-7}$ & $1.4\times10^{-4}$\\
	3,2& 5.591067 & 0.333160 & 3.001563 & $2\times10^{-12}$& 5.591792 & 0.333149 & 3.001656 & $2\times10^{-7}$ & $1.3\times10^{-4}$\\
	3,3& 6.878680 & 0.409886 & 2.439704 & $2\times10^{-12}$& 6.879493 & 0.409869 & 2.439807 & $2\times10^{-7}$ & $1.2\times10^{-4}$\\
	3,4& 8.158826 & 0.486167 & 2.056906 & $2\times10^{-12}$& 8.159719 & 0.486142 & 2.057011 & $2\times10^{-7}$ & $1.1\times10^{-4}$\\
	3,5& 9.433911 & 0.562147 & 1.778895 & $3\times10^{-12}$& 9.434880 & 0.562114 & 1.778998 & $2\times10^{-7}$ & $1.0\times10^{-4}$\\
	3,10& 15.767068 & 0.939526 & 1.064367 & $2\times10^{-11}$& 15.768422 & 0.939456 & 1.064446 & $2\times10^{-7}$ & $8.6\times10^{-5}$\\
\tableline
\end{tabular}
\end{center}
\tablecomments{%
Both stars are scaled to $M=0.6M_{\odot}$ and $R=1.3R_{\oplus}$. %
All modes use $\Gamma_1 = 5/3$.}
\end{table*}

\begin{figure*}
\input{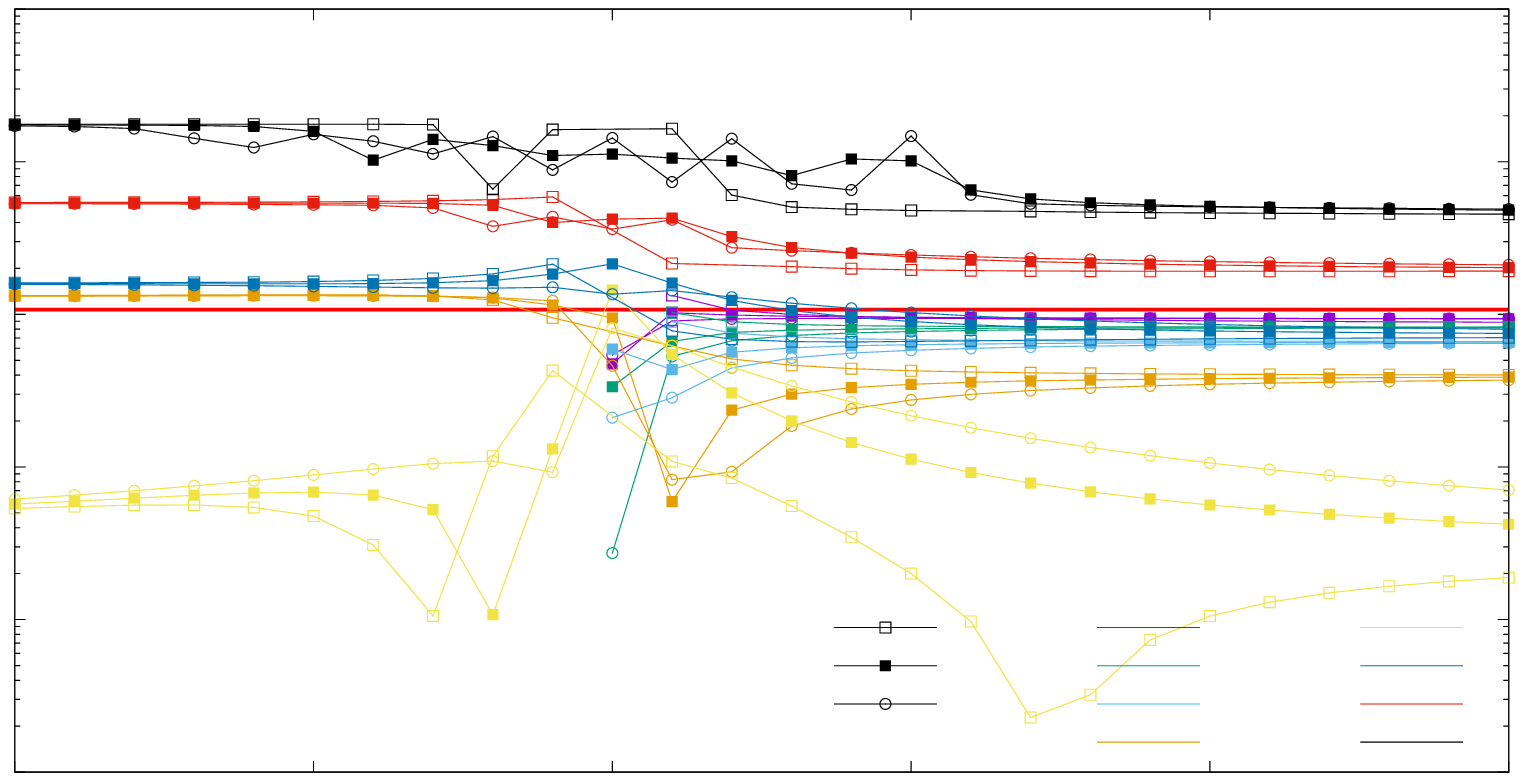}
\caption{Graph of relative differences in $\bar{\omega}$ between models for several polytrope models.  The central 
red line shows $z=1.07\times10^{-4}$.  The dependence on mode type (g- or p-modes) and the dependence on model is 
evident from the graph.  Of particular interest is $n=2.5$, where the difference between Newtonian
and \pn1 frequencies becomes much smaller than $z$, or for $n>3$, where the difference becomes much larger than $z$.
\label{fig:1pnResidualGraph}}
\end{figure*}

\begin{figure*}
\input{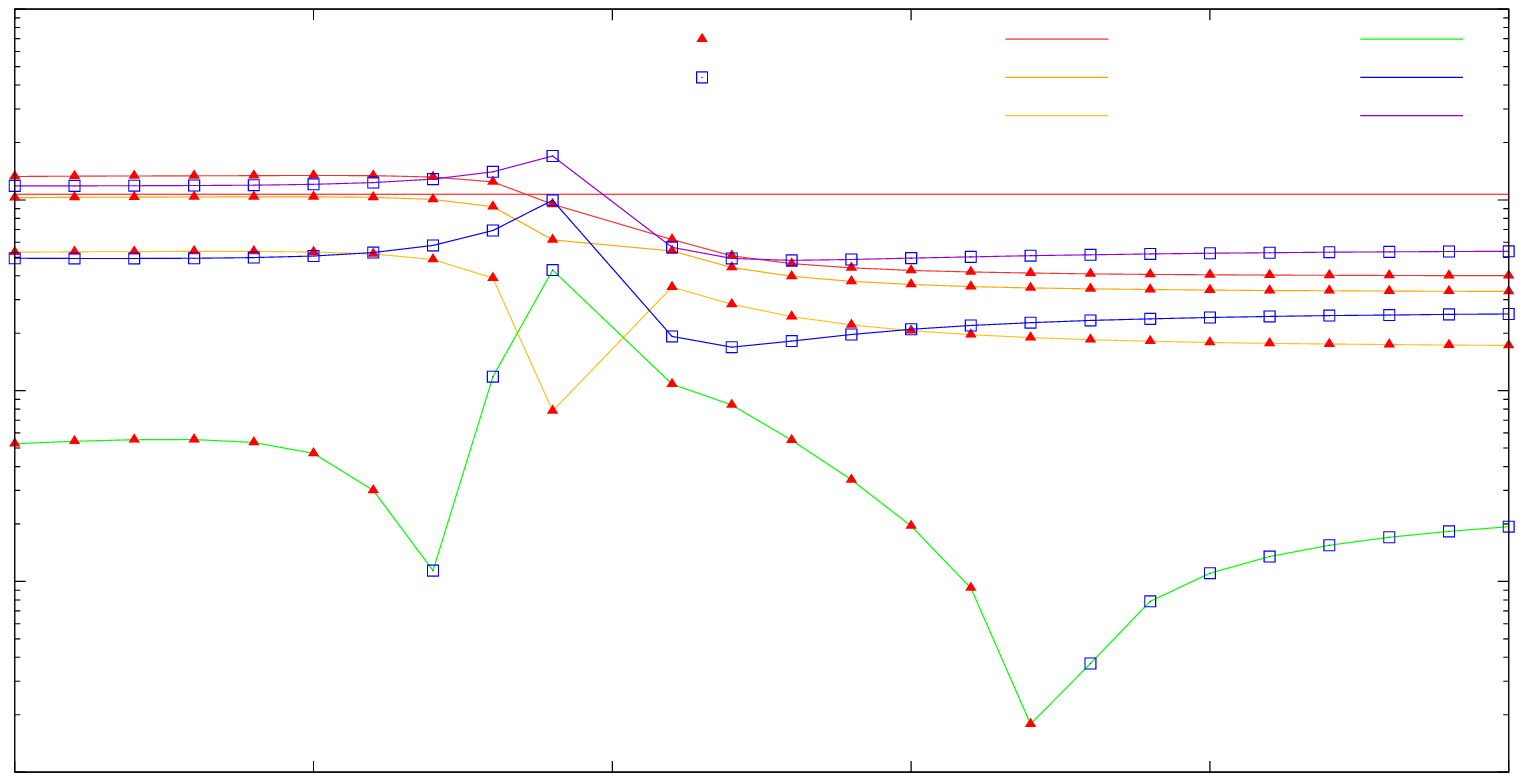}
\caption{Refined graph for many models in the range $n\in[2,3]$, arranged in rough chromatic order with increasing 
$n$, with $\ell=1$ for all modes.  Frequencies with a positive difference (meaning $\bar{\omega}_{\text{\pn1}}$ has 
been blue-shifted) and frequencies with a negative difference (meaning $\bar{\omega}_{\text{\pn1}}$ has been 
red-shifted) are marked with different symbols.  The sign shift occurs between $n=2.5$ and $n=2.7$, suggesting the 
existence of a polytropic index where the difference will be zero for some modes.
\label{fig:1pnResidualGraphZoom}}
\end{figure*}

Results for many more polytrope models are graphed in Figure \ref{fig:1pnResidualGraph}.  A careful look at the 
tables will reveal that between $n=1.5$ and $n=3$ the frequencies change from redshifted to blueshifted.  In Figure 
\ref{fig:1pnResidualGraphZoom} more results are graphed in the range $n=2$ to $n=3$, showing this shift occurs 
around $n=2.6$.

For the CHWD++ models, the choice of $y_0$ fixes $M$ and $R$ according to a mass-radius relation, and therefore it 
is not possible to freely scale $z$ to match between the Newtonian and \pn1 models.  We therefore choose to match 
them according to $M$.  In Table \ref{tab:CHWD++}, we list eigenfrequencies calculated on a CHWD++ background matched 
to $M = 0.607158 M_{\odot}$.  This table records the dimensionless frequency $\bar{\omega}$, the period $\Pi$ 
(in s), and the RMSR, along with the relative difference as in \eqref{eq:compare}.

\begin{table*}
\begin{center}
\caption{Normal mode frequencies and periods for a CHWD++ with $M=0.60716M_{\odot}.$}
\label{tab:CHWD++}
\begin{tabular}{crrcrrcc}
\tableline
	&\multicolumn{3}{c}{Newtonian CHWD++ (RMSR = $1\times10^{-10}$)}
	&\multicolumn{3}{c}{Post-Newtonian CHWD++ (RMSR = $3\times10^{-13}$)}
	&  \\
	&\multicolumn{3}{c}{ $y_0 = 1.57991$, $z=1.0301\times10^{-4}$}
	&\multicolumn{3}{c}{ $y_0 = 1.581  $, $z=1.0317\times10^{-4}$}
	&  \\
$l,k$
	& \multicolumn{1}{c}{$\bar{\omega}$}
	& \multicolumn{1}{c}{$\Pi$ (s)}
	& RMSR
	& \multicolumn{1}{c}{$\bar{\omega}$}
	& \multicolumn{1}{c}{$\Pi$ (s)}
	& RMSR
	& rel. diff. \\
\hline\hline
	1,-8& 0.0104963 & 1711.4055481 & $6\times10^{-6}$& 0.0105004 & 1708.4024156 & $3\times10^{-5}$ & $3.87\times10^{-4}$\\
	1,-7& 0.0119985 & 1497.1386391 & $6\times10^{-7}$& 0.0120032 & 1494.5113986 & $5\times10^{-7}$ & $3.87\times10^{-4}$\\
	1,-6& 0.0140044 & 1282.7042090 & $6\times10^{-7}$& 0.0140098 & 1280.4531595 & $3\times10^{-6}$ & $3.87\times10^{-4}$\\
	1,-5& 0.0168196 & 1068.0042846 & $2\times10^{-7}$& 0.0168261 & 1066.1299007 & $2\times10^{-7}$ & $3.87\times10^{-4}$\\
	1,-4& 0.0210630 & 852.8448434 & $1\times10^{-7}$& 0.0210711 & 851.3479456 & $1\times10^{-7}$ & $3.87\times10^{-4}$\\
	1,-3& 0.0282100 & 636.7745033 & $4\times10^{-8}$& 0.0282210 & 635.6567139 & $9\times10^{-8}$ & $3.88\times10^{-4}$\\
	1,-2& 0.0429319 & 418.4169279 & $1\times10^{-8}$& 0.0429486 & 417.6823108 & $7\times10^{-8}$ & $3.88\times10^{-4}$\\
	1,-1& 0.0945682 & 189.9521474 & $1\times10^{-8}$& 0.0946049 & 189.6186247 & $7\times10^{-8}$ & $3.88\times10^{-4}$\\
\tableline
	2,-8& 0.0181734 & 988.4458576 & $2\times10^{-6}$& 0.0181804 & 986.7117546 & $2\times10^{-5}$ & $3.87\times10^{-4}$\\
	2,-7& 0.0207742 & 864.6997279 & $6\times10^{-7}$& 0.0207822 & 863.1826760 & $5\times10^{-7}$ & $3.87\times10^{-4}$\\
	2,-6& 0.0242469 & 740.8560180 & $6\times10^{-7}$& 0.0242562 & 739.5561899 & $5\times10^{-6}$ & $3.87\times10^{-4}$\\
	2,-5& 0.0291209 & 616.8580678 & $2\times10^{-7}$& 0.0291321 & 615.7757375 & $2\times10^{-7}$ & $3.87\times10^{-4}$\\
	2,-4& 0.0364670 & 492.5937394 & $2\times10^{-7}$& 0.0364812 & 491.7293805 & $1\times10^{-7}$ & $3.87\times10^{-4}$\\
	2,-3& 0.0488399 & 367.8022592 & $4\times10^{-8}$& 0.0488588 & 367.1568083 & $9\times10^{-8}$ & $3.87\times10^{-4}$\\
	2,-2& 0.0743247 & 241.6886386 & $2\times10^{-8}$& 0.0743535 & 241.2644417 & $7\times10^{-8}$ & $3.87\times10^{-4}$\\
	2,-1& 0.1637001 & 109.7338397 & $1\times10^{-8}$& 0.1637635 & 109.5412449 & $7\times10^{-8}$ & $3.87\times10^{-4}$\\
\tableline
	3,-8& 0.0256886 & 699.2757612 & $1\times10^{-6}$& 0.0256986 & 698.0493223 & $6\times10^{-6}$ & $3.86\times10^{-4}$\\
	3,-7& 0.0293645 & 611.7396636 & $8\times10^{-7}$& 0.0293759 & 610.6667318 & $5\times10^{-7}$ & $3.86\times10^{-4}$\\
	3,-6& 0.0342726 & 524.1335104 & $5\times10^{-7}$& 0.0342859 & 523.2142080 & $2\times10^{-6}$ & $3.86\times10^{-4}$\\
	3,-5& 0.0411612 & 436.4171049 & $2\times10^{-7}$& 0.0411771 & 435.6516259 & $2\times10^{-7}$ & $3.86\times10^{-4}$\\
	3,-4& 0.0515434 & 348.5109838 & $3\times10^{-7}$& 0.0515633 & 347.8996633 & $1\times10^{-7}$ & $3.86\times10^{-4}$\\
	3,-3& 0.0690290 & 260.2305123 & $4\times10^{-8}$& 0.0690556 & 259.7740126 & $8\times10^{-8}$ & $3.86\times10^{-4}$\\
	3,-2& 0.1050413 & 171.0130636 & $2\times10^{-8}$& 0.1050819 & 170.7130437 & $7\times10^{-8}$ & $3.87\times10^{-4}$\\
	3,-1& 0.2313130 & 77.6585806 & $1\times10^{-8}$& 0.2314024 & 77.5223589 & $7\times10^{-8}$ & $3.86\times10^{-4}$\\
\tableline
\end{tabular}
\end{center}
\tablecomments{%
All modes use $\Gamma_1$ calculated from background equation of state.}
\end{table*}

\section{Discussion}
\label{sec:discussion}

The results presented in Section \ref{sec:results} generally confirm that GR will lead to measurable changes at the 
\pn1 level in periods derived from space-based photometry for compact stellar objects of similar mass and radius 
as WDs.  This change, represented by the relative difference, is roughly of the same order of magnitude as the 
gravitational surface redshift $z$.  Because the effect is measurable, the full precision of the photometric data 
from K2 and TESS cannot be used to fit asteroseismic models of stars unless GR is first included in both the model 
and perturbation equations.  We recommend the \pn1 formalism as the simplest way to do this, and our set of equations 
\eqref{eq:1pnSystem} might conveniently be added to existing asteroseismology codes.

The neglect of GR is not the largest source of error in normal mode calculations for WDs.  Other factors besides 
\pn1 gravitational fields, such as differences in equation of state, composition, opacity, or treatment of 
convection will have larger numerical impacts.  Using the \pn1 equations does not guarantee a full fit to the 
K2 and TESS data.  However, our results indicate that the size of the \pn1 correction is now observationally 
significant.  One of the main purposes of asteroseismology is to solve the inverse problem, using observed mode 
periods to infer underlying stellar parameters, such as total mass, fractional composition, temperature, radius, 
rotation, existence of solid core, etc.  Any purely Newtonian approach to modeling a pulsating WD, fitting 
eigenperiods to the full resolution of K2 or TESS, will give rise to errors in interpreted values of physical 
parameters at a fractional level of $\sim 10^{-4}$.

The results also show that the underlying stellar model is important in determining not only the extent that GR 
changes the frequency, but whether the frequencies are blue- or red-shifted.  The relative differences for polytropes 
undergo a sign change between $n=2$ and $n=3$, which suggests a balancing act of competing effects.  For larger 
$n$, the mass of the star is more centrally condensed, causing $\beta_\ast^2$ to become distinctly larger than $z$ 
in the center, driving up the difference.  At the same time, with increasing $n$ the star becomes more diffuse near 
the surface, and smaller densities lead to a smaller local strength of relativity in the outer layers, driving down 
the difference.  Somewhere between $n=2.5$ and $n=2.7$, the diffuse atmosphere becomes the most important effect 
and the difference drops to nearly zero, whereas later the increasingly compact core becomes the more important 
effect.
\footnote{%
	Our thanks to Kip Thorne for suggesting this possibility.
}
Whether this same balancing act occurs in a real WD is a subject for future study, but the present work indicates 
that the GR correction in a real WD will not be as simple as merely gravitationally redshifting the frequencies.  

\begin{acknowledgments}

We thank J.~J.~Hermes for helpful discussions and pointing out several important observational results.  This paper 
was partially supported by NSF grants PHY-1506182, PHY-1806447, PHY-2110335 and the North Carolina Space Grant 
Graduate Research Fellowship through NASA.  R.~B.~thanks his mother, father, and wife.
\end{acknowledgments}

\bibliography{master.bib}
\bibliographystyle{aasjournal}

\end{document}